\documentclass[preprint]{aastex}
\shorttitle{Speckle Measures with DSSI. IX.}
\shortauthors{Horch et al.}

\begin{document}
\newcommand{\beq}{\begin{equation}}
\newcommand{\eeq}{\end{equation}}
\newcommand{\ea}{{\it et al.\ }}


\title{Observations of Binary Stars with the Differential Speckle
Survey Instrument. IX. Observations of Known and Suspected Binaries, 
and a Partial Survey of Be Stars}

\author{Elliott P. Horch\altaffilmark{1,11},
Gerard T. van Belle\altaffilmark{2},
James W. Davidson, Jr.\altaffilmark{3},
Daryl Willmarth\altaffilmark{4},
Francis C. Fekel\altaffilmark{5},
Matthew Muterspaugh\altaffilmark{6},
Dana I. Casetti-Dinescu\altaffilmark{1},
Frederick W. Hahne\altaffilmark{1},
Nicole M. Granucci\altaffilmark{1},
Catherine Clark\altaffilmark{2,7},
Jennifer G. Winters\altaffilmark{8},
Justin D. Rupert\altaffilmark{1,12},
Samuel A. Weiss\altaffilmark{1},
Nicole M. Colton\altaffilmark{1,13},
Daniel A. Nusdeo\altaffilmark{9}, and
Todd J. Henry\altaffilmark{10}
}

\affil{$^{1}$Department of Physics, 
Southern Connecticut State University,
501 Crescent Street, New Haven, CT 06515, USA}

\affil{$^{2}$Lowell Observatory, 1400 West Mars Hill Road, Flagstaff, AZ 86001, USA}

\affil{$^{3}$Department of Astronomy, University of Virginia, P.O. Box 400325,
530 McCormick Road Charlottesville, VA 22904, USA}

\affil{$^{4}$NSF's National Optical-Infrared Research Laboratory, 950 N. Cherry Avenue, Tucson, AZ 85719, USA}

\affil{$^{5}$Center of Excellence in Information
Systems, Tennessee State University, 3500 John A. Merritt
Boulevard, Box 9501, Nashville, TN 37209, USA}

\affil{$^{6}$Columbia State Community College, 1665 Hampshire Pike, Columbia, TN 38401, USA}

\affil{$^{7}$Department of Physics and Astronomy, Northern Arizona University, Flagstaff, AZ 86001, USA}

\affil{$^{8}$Center for Astrophysics, Harvard \& Smithsonian, 60 Garden Street, Cambridge, MA 02138, USA}

\affil{$^{9}$Department of Physics and Astronomy, Georgia State University, 25 Park Place, Atlanta, GA 30302, USA}

\affil{$^{10}$RECONS Institute, Chambersburg, PA 17201, USA}

\email{horche2@southernct.edu, gerard@lowell.edu, jimmy@virginia.edu, 
willmart@noao.edu, fekel@evans.tsuniv.edu, astroprofm@gmail.com,
danacasetti@gmail.com,
hahnef1@southernct.edu,
granucci.nicole@gmail.com,
catclark@lowell.edu,
jennifer.winters@cfa.harvard.edu,
jurupert@yahoo.com,
weisss4@southernct.edu, 
nicole.colton@colostate.edu,
nusdeo@astro.gsu.edu, thenry@astro.gsu.edu
}

\altaffiltext{11}{Adjunct Astronomer, Lowell Observatory, 1400 West Mars Hill Road, Flagstaff, AZ 86001 USA}
\altaffiltext{12}{Current Address: MDM Observatory, c/o NSF's National Optical-Infrared Research Laboratory, 
950 N. Cherry Ave, Tucson, AZ 85719 USA}
\altaffiltext{13}{Current Address: Department of Physics, Colorado State University, Fort Collins, CO 80523 USA}

\begin{abstract}

We report 370 measures of 170 components of binary and multiple star systems, obtained from
speckle imaging observations made with the Differential Speckle Survey Instrument at Lowell 
Observatory's Discovery Channel Telescope in 2015 through 2017. Of the systems studied, 147 are binary stars, 
10 are seen as triple systems, and 1 quadruple system is measured.  
Seventy-six high-quality non-detections and fifteen newly resolved components are presented in our observations.
The uncertainty in relative astrometry appears to be similar to our previous work at Lowell, namely
linear measurement uncertainties of approximately 2 mas, and the relative photometry appears to be
uncertain at the 0.1 to 0.15 magnitude level. Using these measures and those in the literature, we
calculate six new visual orbits, including one for the Be star 66 Oph, and two combined 
spectroscopic-visual orbits. The latter two orbits, which are for HD 22451 (YSC 127) and HD 185501 (YSC 135), 
yield individual masses of the components at the level of 2 percent or better, and independent distance measures
that in one case agrees with the value found in the {\it Gaia} DR2, and in the other disagrees at the 2-$\sigma$
level. We find that HD 22451 consists
of an F6V+F7V pair with orbital period of $2401.1 \pm 3.2$ days and masses of 
$1.342 \pm 0.029$ and $1.236 \pm 0.026$ $ M_{\odot}$. For HD 185501, both stars are 
G5 dwarfs that orbit one another with a period of  $433.94 \pm 0.15$ days, and the masses are $0.898 \pm 0.012$ 
and $0.876 \pm 0.012$ $ M_{\odot}$. We discuss the details of both the new discoveries and the 
orbit objects.

\end{abstract}

\keywords{Binary stars: Visual binary stars --- Binary stars: Spectroscopic binary stars ---
Binary stars: Interferometric binary stars --- Stars: B(e) Stars --- Astronomical techniques: Interferometry --- 
Astronomical techniques: Photometry --- Astronomical techniques: Spectroscopy}

\section{Introduction}

There continues to be substantial interest in speckle imaging as a tool for
both stellar and exoplanet science. New speckle instruments
are now resident at the WIYN telescope, as well as both Gemini-North and Gemini-South;
all three of these instruments are based on the design of the Differential Speckle 
Survey Instrument (DSSI; Horch \ea 2009), which is currently resident at 
Lowell Observatory's Discovery Channel Telescope (LDT, formerly DCT). All of these speckle instruments 
take speckle observations in two wavelengths simultaneously, thereby giving color information
for every target observed. This has led to robust observing programs of well-known
binary stars in order to determine orbits and stellar masses ({\it e.g.\ }Horch
 \ea 2017, Horch \ea 2019), surveys of binaries to search for 
tertiary companions in order to differentiate between star formation theories (Tokovinin and Horch, 2016), 
surveys of nearby late-type dwarfs to learn more about multiplicity rates for the
K and M spectral types (van Belle \ea 2018; Clark \ea 2019; Nusdeo, 2018), and 
searches for stellar companions to exoplanet host stars to provide a better
understanding of what conditions produce stable planetary
systems (Horch \ea 2014; Matson \ea 2018; Winters \ea 2019).

Our speckle work at the LDT with DSSI began in 2014, with first results 
reported in Horch \ea (2015). The current paper represents the second installment
of this effort. In the first paper, the objects observed were primarily 
{\it Hipparcos} doubles and suspected double stars (ESA 1995), and stars listed in the Geneva-Copenhagen
spectroscopic survey as double-lined spectroscopic binaries (Nordstr\"{o}m \ea 2004). Starting in 2015, we
began to use the majority of the time awarded to make progress on the surveys of K and M dwarfs in the 
Solar neighborhood; the observations and analysis of the stars in these surveys are 
ongoing. The remainder of the time was used to obtain observations of the few targets
that had not been observed from the earlier observing list (that is, the {\it Hipparcos} and 
Geneva-Copenhagen stars), as well as
to obtain observations for an initial survey of Be stars, in order to search for previously undetected
stellar companions orbiting these high-mass stars. Binarity is expected to be extremely common 
for early-type main sequence stars as discussed in Duch\^{e}ne and Kraus (2013) and references therein,
and speckle observations would help develop an understanding of the stellar multiplicity rate 
for Be stars, 
relative to B stars without emission lines. In addition, when combined with other measurements that yield insight
into the properties of the emission disk, it can give a better understanding of the interaction between
the companion and the disk in specific cases (Bjorkman \ea 2002; Rividius \ea 2013). It is the
speckle observations of {\it Hipparcos}, Geneva-Copenhagen, and Be stars obtained since the
publication of our last set of measures that we report on in this paper. 

\section{LDT Speckle Observations and Data Reduction}

The speckle observations presented here were taken on several runs beginning in March 2015 and 
ending in May 2017 using the Differential Speckle Survey Instrument (Horch \ea 2009). After a long 
period of time resident at the WIYN Telescope at Kitt Peak National Observatory (2008 through 
2013), and several runs as a visitor instrument at the Gemini-North telescope (July 2012 through 
January 2016),  the instrument has most recently been shared by Gemini-South and Lowell 
Observatory (at the LDT). In order to accommodate scheduling constraints at both 
Gemini-South and at Lowell, the runs for the project described here were often completed 
when a port on the LDT instrument cube became vacated by one of the other instruments
used there. As a result, a different port was used from run to run. Nonetheless, the telescope
delivers the same plate scale to all of the ports on the instrument cube, so the only change 
this presents is the orientation of the image on the chip. With the data we have in hand, we
have so far not identified any systematic differences in scale that are related to any of the ports
used.

The standard procedure for speckle observing at the LDT is to select a 
bright unresolved star within a few degrees of the each science target, usually from the Bright 
Star Catalog (Hoffleit and Jaschek, 1981). This star serves as a ``point source calibrator'' for the
science target; that is, an estimate of the speckle transfer function for the observation of the science
target. If two or more science targets are clustered closely on the sky, a single point source 
is used for the entire group, so that the observing list for the night is divided into blocks, 
each of which has a single point source. Blocks are then ordered in right ascension, and
objects inside each block are also ordered in the same way. This allows for the observations 
to commence at the beginning of each night on or near the meridian, and then for objects to be observed in sequence
at small hour angle when they are near the meridian, and therefore near the minimum airmass.
This regimen not only results in mainly small telescope moves, but is also important because
neither the DSSI nor the LDT have atmospheric dispersion compensation prisms (Risley prisms),
and so large zenith angles, which can degrade the quality of the observation, are avoided if possible.

Most of the stars discussed here are bright enough that a single set of 1000 40-ms speckle frames
was sufficient to obtain a robust result. DSSI is equipped with two Andor iXon 897 
electron-multiplying CCD cameras, and sub-arrays of either 128$\times$128 or 256$\times$256
pixels centered on the star were read out. The two data files
are stored in FITS format. For all of the observations reported here, the filters used
were a 692-nm filter of width 40 nm, and an 880-nm filter of width 50 nm.

The reduction of the data proceeds along the same lines as we have outlined in our previous LDT
work (Horch \ea 2015). Average autocorrelations of both the target observation and the point 
source are formed and Fourier transformed to arrive at the spatial frequency power spectrum in 
each case. Subplanes of the image bispectrum are also calculated in the case of the target star,
following Lohmann \ea (1983). The modulus of the true object Fourier transform is obtained by 
dividing the target power spectrum by that of the point source and taking the square root; the 
phase of the objects Fourier transform is found from the bispectral subplanes using the relaxation
algorithm of Meng \ea (1991). This phase is combined with the modulus estimate, low-pass filtered
with a two-dimensional Gaussian function of width comparable to the diffraction limit, and the result
is inverse-transformed to arrive at the reconstructed image for the target.

The reconstructed images are then studied and if one or more companions are present, the pixel 
coordinates of each are noted. These are then used as input for a fitting routine that finds the best 
fit fringe pattern to the deconvolved binary power spectrum in the Fourier domain, as described by 
Horch \ea (1996). This routine outputs the final separation ($\rho$), position angle ($\theta$) and 
magnitude difference ($\Delta m$) of each detected companion.
If a companion is not detected, then a detection limit curve is generally made,
which, based on statistics of the reconstructed image, attempts to estimate
the magnitude difference of the faintest companion relative to the
primary star that would be detected, as a function of separation. 

%\subsection{Pixel Scale and Orientation}

The pixel scale and orientation were determined using a small set 
of ``calibration binaries;'' that is, binary stars with recent orbits determined 
with the inclusion of long-baseline optical interferometry data. Table 1 shows 
the binaries and the orbit references that were used for this purpose. The
orbital elements listed in those references were used to calculate the 
separation and position angle of the pair at the time of observation at the LDT,
and comparing to the final values from the fringe fits from the speckle data, the pixel scale and 
orientation are derived.

\section{Spectroscopic Observations and Data Reduction}

Beginning with the use of DSSI at the WIYN telescope,
we have been interested in obtaining spectroscopic
observations of the smallest-separation systems that we
have discovered.
Such observations complement the ongoing speckle work and provide 
a means to eventually determine individual masses to high precision.
A short list of suspected close speckle binaries was made, and  
we began spectroscopic observations in 2012 October with the Kitt Peak National
Observatory (KPNO) 0.9~m auxillary coud\'{e} feed telescope and the
coud\'{e} spectrograph, which was originally built for use with the
KPNO 2.1~m telescope. 

The KPNO CCD spectra were calibrated and extracted with standard IRAF tasks,
after which radial velocities were measured with the IRAF task FXCOR
(Fitzpatrick, 1993). Template stars for the cross correlations were from the
list of Scarfe (2010) or from Nidever \ea (2002). A comparison of velocities
from these two sources shows good agreement with a mean difference
(RV$_{\rm Nidever} - RV_{\rm Scarfe}$) of
0.08 $\pm$0.04~km~s$^{-1}$ for 6 stars common to the two lists. A
seventh star, HD 103095, has a difference of 0.51~km~s$^{-1}$ and is
possibly variable in velocity. 

In the case of two small--magnitude-difference systems on our observing list, double-lined spectra were 
quickly detected and these objects have been re-observed a number of times. These
are YSC~127 = HD 22451 = HIP 17033
and YSC~135 = HD 185501 = HIP 96576.
Our KPNO observations, 7 of YSC~127 and 7 of YSC~135,
were obtained with various combinations of KPNO telescopes and instruments,
which are listed in Table~\ref{tbl-instcomb}. The spectra have resolving
powers that range from 16500 to 72000.
The KPNO observations and resulting velocities
of YSC~127 and YSC~135 are listed in Tables~\ref{tbl-22451obs} and
\ref{tbl-185501obs}, respectively.
An extensive number of additional observations were acquired from
2015 April through 2019 April at Fairborn Observatory in
southeast Arizona. During that period we obtained 68 spectra of
YSC~127 and 39 spectra of YSC~135 with the Tennessee State University
2 m Astronomical Spectroscopic Telescope (AST) and a fiber fed echelle
spectrograph (Eaton \& Williamson, 2004). Our detector, a Fairchild
486 CCD, has a 4096 $\times$ 4096 array of 15 $\micron$m pixels and
a resulting wavelength coverage of 3800 to 8600~\AA\ (Fekel \ea 2013).
The AST spectra have a resolving power of 25000 at 6000~\AA\ and
signal-to-noise ratios of about 85 for YSC~127 and 100 for YSC~135.
These signal-to-noise ratios were estimated from the rectified
continuum scatter at about 6000~\AA. 
Eaton \& Williamson (2007) explained the reduction and wavelength calibration of the
raw spectra.

Fekel \ea (2009) have given a general description of the typical reduction 
leading to the velocity determination.
Specifically, for YSC~127 and YSC~135 we used a solar line
list that consists of 168 lines in the spectral region 4920--7100~\AA.
Each unblended line was fitted with a rotational broadening function
(Lacy \& Fekel 2011, Fekel \& Griffin 2011).
Measurement of the lines when the components were
blended consisted of simultaneous fits of the blended components with
two rotational broadening functions. Our unpublished velocities for
several IAU solar-type velocity standards show that our velocities
have a $-$0.6 km~s$^{-1}$ shift relative to the results of Scarfe (2010).
Thus, we have added 0.6~km~s$^{-1}$ to all our velocities. Our AST
observations and velocities for YSC~127 and YSC~135 are listed in
Tables~\ref{tbl-22451obs} and \ref{tbl-185501obs}, respectively.

Our radial velocities from the two observatories are tied to the IAU
standards observed by Scarfe (2010). Thus, there is no significant
zero point shift in the velocities from the two observatories
(Willmarth \ea 2016).

\section{Speckle Results}

Our main body of speckle results is presented in Table 5. The columns give:
(1) the Washington Double Star (WDS) number (Mason \ea 2001)\footnotemark, which also
\footnotetext{{\tt http://www.astro.gsu.edu/wds/}}
gives the right ascension and declination for the object in J2000.0 coordinates;
(2) the Aitken Double Star (ADS) Catalogue number,
or if none, the Bright Star Catalogue ({\it i.e.,\ }Harvard Revised [HR]) number,
or if none, the Henry Draper Catalogue (HD) number,
or if none the Durchmusterung (DM) number of the object;
(3) the Discoverer Designation;
(4) the {\it Hipparcos} Catalogue number;
(5) the Besselian date of the observation;
(6) the position angle ($\theta$) of the
secondary star relative to the primary, with North through East defining the
positive sense of $\theta$; (7) the separation of the two stars ($\rho$), in
arc seconds; (8) the magnitude difference ($\Delta m$)
of the pair in the filter used;
(9) the center wavelength of the filter; and
(10) the full width at half maximum of the filter transmission.
The position angle measures have not been precessed from the dates shown.
Fifteen pairs in the table have no
previous detection of the companion
in the {\it Fourth Catalogue of
Interferometric Measures of Binary Stars} (Hartkopf,
\ea 2001a; hereafter Fourth Interferometric Catalog)\footnotemark; we propose discoverer designations of
LSC (Lowell-Southern Connecticut) 116-130 here. (This continues the collection
of LSC discoveries detailed in Paper VI in this series.)
\footnotetext{{\tt http://www.astro.gsu.edu/wds/int4/}}

To illustrate the properties of the data set overall, we plot in Figure 1(a) the magnitude difference of our measures 
as a function of the separation measured. In Figure 1(b), the magnitude difference is plotted as a function of 
the $V$ magnitude of the system, as it appears in the {\it Hipparcos} Catalogue. In Figure 1(a), we have superimposed 
onto the plot a typical detection limit curve as discussed in Section 3.3. As described further there, this curve is estimated 
by examining noise statistics in annuli of the reconstructed image centered on the primary star. Below the detection 
limit determined at the smallest radius in the calculation (0.1 arcsec for all observations here), 
we assume a linear fall-off in the detection
limit down to zero at the separation corresponding to the Rayleigh criterion. We see that 
the envelope of all the components detected matches 
well with the detection limit estimate that is drawn in Figure 1(a), above a separation of 0.1 arc seconds. Below that, 
as the data points indicate, we are in fact able to measure the separation of components with decreasing sensitivity in magnitude
difference as the separation decreases, roughly along a linear trend as indicated by the black line in the figure. 
(The measurements falling in this part of the diagram are generally known to be binary from previous, often spectroscopic,
observations.) In Figure 1(b), we see that most of the stars listed in Table 5 have magnitudes between $V = $ 5 and 11.

\subsection{Relative Astrometry}

To gauge the uncertainties that should be associated with the measures in Table 5, we first study the repeatability of the
results for the measures obtained in the two filters during the same observation. Differences in separation and in position angle
between the two channels are shown in Figures 2(a) and 2(b) respectively. For position angle, an average difference of 
$0.13 \pm 0.11$ degrees is obtained from the entire data set, with a standard deviation of $1.46 \pm 0.08$ degrees. However, 
as is well known, the position angle uncertainty grows at smaller separation due to the fact that the same linear measurement
error will subtend a larger position angle; if only separations larger than 0.05 arc seconds are considered, the mean difference
becomes $0.04 \pm 0.06$ degrees with a standard deviation of $0.82 \pm 0.04$. In terms of the separation measures, a slight 
offset between the channels is noted at the 2-$\sigma$ level; 
the average difference is $0.4 \pm 0.2$ mas, and although small, may indicate that
a better system of scale calibration should be used in the future. The standard deviation of these differences is $2.4 \pm 0.1$ mas. 
If only separations larger than 0.05 are included, the mean
remains unchanged and the standard deviation is slightly lower, $2.3 \pm 0.1$ mas. 

The standard deviation numbers for both position angle and separation are very similar to the previous large group of measures
from the LDT reported in Horch \ea (2015). As discussed there, because these are derived from the difference between two independent 
measures that presumably have similar random uncertainties, we may conclude that the 1-$\sigma$ internal precision of the data set 
in Table 5 is given by these standard deviations divided by $\sqrt{2}$. Thus, subject to the caveat that the position angle 
uncertainty is a function of separation as discussed above, we may conclude that the internal precision of the measures in Table 5
is generally $0.58 \pm 0.03$ degrees in position angle and $1.6 \pm 0.1$ mas in separation over the magnitude range represented
by the sample.

Because we have observed a number of well-known binary stars in the data set presented, 
we also have the opportunity to compare to the ephemeris
positions of those stars with orbits in the literature (excluding, of course, those objects used in the determination of the scale
and orientation for each run). A list of such objects with orbits of Grade 2 or better in the {\it Sixth Catalog of Orbits of Visual 
Binary Stars} (Hartkopf \ea 2001b; hereafter Sixth Orbit Catalog\footnotemark) are shown in Table 6.
\footnotetext{{\tt http://www.astro.gsu.edu/wds/orb6}}
Also included there are the ephemeris position angles and separations, with uncertainties as calculated from the uncertainties in
the orbital elements, and the observed minus ephemeris residuals when comparing to the observed values in Table 5. In Figure
3, we plot these results. To make the comparison, we have averaged the astrometric results in both channels of the instrument, 
whereupon the precision obtained should be that of a single measure divided by $\sqrt{2}$. As determined by the comparison
between the channels given above, the precision of a single measure is 1.2 mas in separation and 0.42 degrees in position angle.
We show in Figure 3 these values, using $\tan(1.2/\rho)$ (where $\rho$ is in mas) to convert the linear measurement 
precision into an angular one; for
the mean separation of this group of objects, that results in a median angular precision of 0.20 degrees. Together, the plots
indicate that there are no obvious offsets in position angle or separation, and that the accuracy of these measures relative to 
the orbital ephemerides is comparable to the internal precision.

As a further check, we also found that six of our systems with
separations greater than one arc second had astrometry for both components in the {\it Gaia} DR2 (Gaia Collaboration, 2018). 
In comparing our measures
to the position angles and separations implied by {\it Gaia}, we find that the average position angle difference was $0.44 \pm 0.52$
degrees and the average separation difference was $8.3 \pm 16.7$ mas. Although the sample is small and there may be some
relative motion of the system in between the mean {\it Gaia} epoch of observation and our measures, this nonetheless gives
some further confidence that the astrometry has been properly calibrated.

\subsection{Relative Photometry}

We turn next to the photometric precision and accuracy of the measures in Table 5. This is harder to judge 
than in the case of relative astrometry because the best 
space-based measures of magnitude differences for the objects in Table 5 
in the literature are usually still those obtained by {\it Hipparcos}; the DR2 of {\it Gaia} does not
contain much information for the objects we have observed due to their mostly sub-arcsecond separations. The {\it Hipparcos}
results are in the so-called $H_{p}$ filter, which has center wavelength of 511 nm and a width $\Delta \lambda$ of 800 nm. This
is not a particularly good match for the results in Table 5, which are considerably redder, and with narrower bandpasses.

Despite this deficiency, we nonetheless show in Figure 4 a comparison between the $\Delta H_{p}$ value appearing in the 
{\it Hipparcos} Catalogue and the magnitude differences appearing for each filter in Table 5. To make the comparison as 
valid as possible, we have limited the range of $B-V$ color for the systems we have plotted to $0.0 < B-V < 0.6$ and excluded
systems that appear to be giants based on their placement in the color-magnitude diagram. 
The horizontal axis of Figure 4(a) is the seeing 
estimate for the observation multiplied by the separation of the components; in previous papers, we have argued that this 
combination is a measure of the ``isoplanicity'' of the observation. That is, it is related to the degree to which the primary and 
secondary star will
have similar speckle patterns. This is because the seeing is inversely proportional to the Fried parameter, while the size of the 
isoplanatic angle is linearly proportional to it. In dividing the separation by the size of the isoplanatic angle, one obtains
a ratio $q$, with $q < 1$ if the secondary falls inside the the isoplanatic region of the primary and 
$q > 1$ if it does not. By multiplying seeing times separation, a parameter $q^{\prime}$ carrying units of 
arcseconds squared is instead obtained, but the value is proportional to $q$. Therefore, one expects that there will be
a particular value below which the observation is isoplanatic, whereas above it, the magnitude difference obtained 
will be affected by the decreased correlation between primary and secondary speckle patterns.

Figure 4(a) shows a result similar to previous papers in this series; the difference between space-based relative photometry 
of the pairs and the speckle result clusters near zero if the seeing times separation is less than approximately 0.6 arcsec$^2$,
but trends upward for larger values of this parameter. This is understandable given that, as the value increases and therefore 
the speckle patterns of primary and secondary are less and less identical, the derived magnitude difference from the speckle
observation is larger and larger as the speckles fail to correlate at the same vector separation in the autocorrelation. 
Because of this, for observations shown in Table 5 with $q^{\prime} > 0.6$ arcsec$^{2}$, the magnitude difference is shown
as less than the value obtained in the fit.
In Figure 4(b), we show a plot of the $\Delta H_{p}$ values as a function of the magnitude difference at 692 nm
for observations with seeing times separation of less than 0.6 arcsec$^2$. Here the data are essentially linear, with standard 
deviation of 0.21 magnitudes if we consider $0.1 < \Delta H_{p} < 4$; some of this scatter is due to the uncertainty in the 
{\it Hipparcos} values themselves; if we 
subtract the average value of the stated uncertainty in $\Delta H_{p} $ in quadrature, which is 0.16 magnitudes, 
then the uncertainty left over and presumably
due to the speckle measurement is 0.13 magnitudes. This is slightly larger than in some previous papers in this series, particularly
measures taken at the WIYN telescope, and if the program of speckle observations continues at the LDT, then this will
have to be studied in more detail. Perhaps further data releases of the {\it Gaia} Collaboration could help resolve this issue.

\subsection{Nondetections}

The {\it Hipparcos} suspected doubles and the Be stars observed in the data set presented here were not known to have companions.
Likewise, for the double-lined spectroscopic binaries observed from the Geneva Copenhagen Catalog, it was not known if 
the resolution limit would permit the detection of the secondary at the LDT. 
In general, we search for companions visually using reconstructed images made from the data. If none is clearly identified, then 
we use the following methodology to establish the detection limit as a function of separation from the primary star. We
identify pixels in annuli centered on the primary star with center width in increments of 0.1 arc seconds. We find local maxima
within each annulus, and compute the average and standard deviation of the peak values of these features. The detection limit for
that annulus is then the average peak value plus five times the standard deviation, converted to a magnitude difference. For
the purpose of the robust discovery of new companions, we require conservatively that the detection limit at the diffraction
limit corresponds to a zero magnitude difference. Once 
values of the detection limit are obtained for the entire set of annuli, a cubic spline routine is used to interpolate between these
values in order to obtain a continuous curve. Examples of curves obtained for both an unresolved star and a companion
previously unknown but appearing in Table 5 are shown in Figure 5.

In all, seventy-six stars in the data set were observed in good conditions and were found to be unresolved, to the limit of detection
of the DSSI camera at the LDT. These are listed in Table 7. We provide there an estimate of the magnitude difference that
represents a 5-$\sigma$ detection at both 0.2 and 1.0 arc seconds from the primary, in both the 692- and 880-nm filters. The 
detection limit curve may be roughly reconstructed from these numbers, as it is usually fairly linear between 0.2 and 1.0 arc seconds,
and again roughly linear (though with different slope) between the diffraction limit and 0.2 arc seconds. 

\subsection{Comments on the Newly Discovered Systems}

We give some further information regarding the fifteen new discoveries here; we take basic stellar data
for each system from SIMBAD (Wenger \ea 2000)\footnotemark.
\footnotetext{{\tt http://simbad.u-strasbg.fr/simbad/}}

\noindent
{\bf LSC 116Aa,Ab = HIP 32887.} Our discovery of a small-separation component of this previously known 
binary MLR 688 reveals this system to be a heirarchical triple system. The composite spectral type is F8, 
and the system is at a distance of approximately 114 pc based on the {\it Hipparcos} parallax result. (No {\it Gaia}
result is available at present.) If 
the new component detected here is indeed a physical member of the system, the current projected separation
is 7 AU. If this is comparable to the semi-major axis, the period would likely be in the
range of 10 to 15 years. 

\noindent
% GC, HD 57863, [Fe/H]=-0.03, mratio= $0.800 \pm 0.050$
%F7V, $\pi = 13.6682 \pm 0.1198$ ({\it Gaia} DR2)
{\bf LSC 117 = HIP 35775.} The component of the F7V system measured here may well be the spectroscopic component
listed in the Geneva-Copenhagen catalog, given the small separation and the {\it Gaia} DR2 distance of 73 pc (Gaia
Collaboration, 2018). Those data
suggest a projected physical separation of 1.8 AU, and a period on the order of a year. The mass ratio is given in the 
Geneva-Copenhagen Catalog as $0.800 \pm 0.05$, and the iron abundance is near solar, 
which is consistent with the relative photometry we have determined
here for a mid-F primary and later-F secondary star.

\noindent
%HIP (but this is a third component)
%F5, $\pi = 2.6493 \pm 0.0721$ ({\it Gaia} DR2.
{\bf LSC 118Aa,Ab = HIP 37657.} This star has a composite F5 spectrum according to information in SIMBAD, and was
also discovered to be binary by the {\it Hipparcos} satellite, where it is listed in the {\it Hipparcos} Catalogue as HDS 1092. 
We detect a small-separation companion to the 
primary star of that system. However, the {\it Gaia} DR2 parallax is only 2.65 mas, and the apparent V magnitude is 8.79;
thus the primary star would appear to be evolved, as this implies an absolute magnitude of $+0.91$, which is too bright for an
F5 dwarf.
This comports with the magnitude differences for both the close and wider companions of over 2 magnitudes. 
Those stars may yet be on or near the main sequence. 

\noindent
%HIPS
%K0, $\pi = 3.1838  \pm 0.0529$ ({\it Gaia} DR2
{\bf LSC 119 = HIP 43519.} The primary of this pair appears to be a K0 giant, given the distance (over 300 pc according to 
the DR2), and apparent $V$ magnitude of 7.872. The component we detect is over 4 magnitudes fainter and at a separation of
0.3 arcseconds, leading to a projected physical separation of about 100 AU. Thus, it is almost certainly not the 
known companion OCC 382. This could be an optical double.

\noindent
%HIPS
%K0-1III, $\pi = 7.4067 \pm 0.0573$ ({\it Gaia} DR2
{\bf LSC 120 = HIP 43810.} Suspected as a binary by {\it Hipparcos}, the primary here is again an early-K giant, 
but this system lies much closer to the solar system at only 135 pc. The separation of the component we 
detect is large (0.7 arcsec) and the
magnitude difference is again over 4 magnitudes, so again the physical association cannot be plausibly established at this point.
The system was seen as unresolved by Mason \ea (1999), probably due to the large magnitude difference.

\noindent
%HIPS
%F5, 6.6889 [0.5051] ({\it Gaia} DR2
{\bf LSC 121 = HIP 44908.} An F5 pair at a distance of 150 pc, this system has a small magnitude difference. If we assume
two F5V stars with a semi-major axis of 15 AU separation (which is the current projected physical separation
given the angular separation of 0.1 arcsec), that would imply an orbital period of less than 40 years, so follow-up observations 
of the system would be helpful in the coming years to confirm orbital motion.

\noindent
%GC, HD 83190, [Fe/H]=-0.08, mratio= $0.549$
%F8, 6.80 [0.71] (HIP revised).
{\bf LSC 122Ba,Bb = HIP 47196.} The secondary of the known binary STF 1372 is revealed here to be a small-separation
binary itself; this is the only example in our list of new discoveries of an A-BC architecture among the five triples. This 
small-separation component may be the spectroscopic one mentioned in the Geneva-Copenhagen catalog. The 
mass-ratio given there is 0.549, but without an uncertainty, and the [Fe/H] value is listed as $-$0.08. There is no DR2 
parallax, but the revised {\it Hipparcos} value is $6.80 \pm 0.71$ mas, indicating a projected physical separation of 1.8 AU,
and if we use this as an estimate of the semi-major axis, the period would be on the order of 2 years. However, the 
magnitude difference between the Ba and Bb components would be on the order of a magnitude given the data in Table
5, and that would most likely give a mass ratio for an early-G and late-G dwarf of closer to 0.8. An orbit due to Alzner
(2005) exists for the AB system; a reanalysis of earlier speckle
observations that already appear in the Fourth Interferometric Catalog might yield more astrometry of this pair, and 
if so, a combined orbit for both the wide and small-separation pairs could be within reach very soon.

\noindent
%HIPS
%A5, 5.0389 [0.2685] ({\it Gaia} DR2
{\bf LSC 123 = HIP 48044.} This star, suspected of having a companion by {\it Hipparcos}, is at a distance of 200 pc, 
so that the current projected separation is then approximately 18 AU. The spectral type is listed in SIMBAD as A5.

\noindent
%HIP (but this is a third component)
%G5, 4.69 [0.78] (HIP revised)
{\bf LSC 124Aa,Ab = HIP 49495.} A sub-arcsecond component to this G5 system has been known for decades. We detect that
companion, now at 0.35 arcsec from the primary, but also find a very small-separation component. The DR2 does not
give a parallax, but the {\it Hipparcos} revised value is $4.69 \pm 0.69$ mas, thus the projected separation is about 5.5 AU.

\noindent
%HIPS
%F0, 5.4510 [0.6343] ({\it Gaia} DR2
{\bf LSC 125 = HIP 49677.} This F0 star was suspected double by the {\it Hipparcos} satellite. We find a component with
magnitude difference of about 2.5 at a separation of 0.38 arcsec. The {\it Gaia} DR2 parallax is $5.4510 \pm 0.6343$ mas.
The system was reported as unresolved by Mason \ea (1999).

\noindent
%HIPS
%G5, 8.91 [1.64] (HIP revised)
{\bf LSC 126 = HIP 52932.} This star is another {\it Hipparcos} suspected double that is resolved here for the first time, although
the separation is over 1 arcsec. The magnitude difference is large, and this is the reason the system was not discovered and
measured by the satellite.
The distance is approximately 112 parsecs.

\noindent
%GC, HD 94686, [Fe/H]=-0.13, mratio= $0.982$
%F8V+F8V, 14.60 [0.60] (HIP revised)
{\bf LSC 127Aa,Ab = HIP 53709.} This star is listed as a double-lined spectroscopic binary star in the Geneva-Copenhagen 
catalog; the mass ratio there is listed as 0.982 (without an uncertainty). A 3-arcsecond companion was already known and
was outside our field of view. However, it
is unlikely to be the spectroscopic component, given the distance of 68 pc; this translates into a projected separation for that
component of 200 AU. In contrast, the small-separation component we find has projected separation of approximately 1.7 AU.
Given the primary spectral type of F8V and a magnitude difference of about 1.5, the secondary, if bound, is likely to be
in the late-G range. 

\noindent
%HIPS
%F5, 7.7238 [0.2526] ({\it Gaia} DR2
{\bf LSC 128 = HIP 53969.} We confirm a faint component at a separation of 0.33 arcsec to this {\it Hipparcos} suspected 
double star. The distance is approximately 130 pc; if bound to the primary, the projected separation is 43 AU.

\noindent
% HR 7266 (serendipidous)
%F0III-IV, 22.9563 [0.2397] ({\it Gaia} DR2
{\bf LSC 129 = HIP 94068.} This is the first of two seredipitous discoveries where we observed the star as a bright
unresolved calibration star, in this case, HR 7266. The companion has a modest magnitude difference and is at a separation
of 0.07 arcsec. Given the relatively small distance to the system of 43.5 pc, this should be an interesting pair to follow
up on in the coming years. Data in SIMBAD suggest that the primary star is an evolved star of spectral type F0.

\noindent
% HR 8217 (serendipitous)
%A1V, 17.0458 [0.1435] ({\it Gaia} DR2
{\bf LSC 130 = HIP 105966.} This is another seredipitous discovery, but in this case the companion found is faint and at a separation
of 0.36 arcsec. The primary star has spectral type A1V, and the secondary would appear to be much fainter and redder;
the photometry in Table 5 would suggest perhaps something in the late-G or early-K range. Two previous speckle observations,
by McAlister \ea (1987) and DeRosa \ea (2011), did not detect the secondary.

\section{New Orbital Elements}

The astrometry in Table 5, together with the radial velocities reported in Section 3 and 
previous measures in the literature, gives us the opportunity to
compute or revise orbital elements for several binaries. These orbits fall into two categories with the larger group
being standard visual orbits. Two objects have also been studied by three of us (D.W., F.F., and M.M.) in spectroscopic
observations over the last several years. In those cases, we are able to present visual+spectroscopic orbital
elements, including individual masses and independent distances.

\subsection{Visual Orbits}

Objects for which we calculate classical visual orbits are shown in Table 8. The method used is that of MacKnight and Horch
(2004), which is a two-step process to deriving final orbital elements. After obtaining the previously available astrometry
of the system from the Fourth Interferometric Catalog, a grid search is performed to identify the orbital
elements that minimize the reduced-$\chi^{2}$ when comparing ephemeris position to the actual measures. The second
step is to use a downhill simplex algorithm to refine those orbital elements to the absolute minimum reduced-$\chi^{2}$.
Uncertainties in the orbital elements are estimated by adding random deviates in $\rho$ and $\theta$ of a typical value
for large-telescope speckle observations (2.5 mas) and recomputing the orbit. 
This is done many times and gives a distribution for each orbital element.
The uncertainties
of each are estimated to be the standard deviation of its distribution. The orbits are shown in Figure 6, and comments 
on each system follow. Further information regarding the calculation of visual orbits is given in Horch (2013) and
references therein.

\noindent
{\bf BU 314AB.} This mid-F star sits at a distance of 40 pc and had a previous Grade 3 orbit in the Sixth Orbit Catalog due to
S\"{o}derhjelm (1999). However, since the calculation of that orbit, there has been a sequence of very good quality speckle observations
leading up to the present that permitted a redetermination of the orbital elements. This was needed here, because the 
system was pressed into service as a scale calibration object for the February 2016 run due to the lack of other suitable 
observations. In combination with the {\it Hipparcos} revised parallax (no {\it Gaia} result is available), our orbit gives a total 
mass of $1.85 \pm 0.14$ $M_{\odot}$.

\noindent
{\bf HDS 1199.} The spectral type for this system is listed in SIMBAD as K7V+M0/2V, and the {\it Gaia} DR2 parallax is
$26.4436 \pm 0.5831$ mas. No previous orbit exists in the Sixth Orbit Catalog. Using the observations 
presented in 
Table 5 together with those in the literature, the data so far span approximately 25 years of the 133-year period, and 
trace out approximately one-quarter of the orbit in terms of the position angle coverage, but nonetheless we determine 
a semi-major 
axis with only  3\% uncertainty. This, together with the parallax and period, yields a mass sum of $2.53 \pm 0.38$ 
$M_{\odot}$, much higher than expected for two late-type dwarfs. However, Parihar \ea (2009) find the star is probably 
an eclipsing binary and as our orbit does not eclipse, a third star must be present in the system. Parihar \ea (2009) 
also find that the spectrum of the system exhibits $H_{\alpha}$ emission. 
We also note that the magnitude differences for the pair measured so far are in the range
of about 2.5, again much larger than expected for a late-K plus early-M star.

\noindent
{\bf COU 1258.} Ours is the first attempt to calculate the orbit of this system, which is of mid-F spectral type and at a distance
of approximately 150 pc. The mass sum we derive at this stage is $2.67 \pm 1.51$ $M_{\odot}$, which is very uncertain but nonetheless
consistent with this spectral type and the small magnitude difference that we have measured. The system was reported as 
unresolved in a 2008 observation by Gili and Prieur (2012); our orbit predicts a separation of 94 mas at the time of their
observation. That observation was done with a 0.7-m telescope using a 570-nm filter, so the system would have been below the 
diffraction limit of 0.21 arcsec.

\noindent
{\bf HDS 1542AB.} SIMBAD lists the spectral type of this system as M1V, and we measure a modest magnitude difference.
The distance is about 40 pc, thus we derive a mass of $\sim$0.7 $M_{\odot}$ with large uncertainty, consistent at this early 
stage with the spectral information. If on the other hand one uses the {\it Gaia} DR2 parallax for the primary (no value exists
in the data release for the secondary), then the distance is 31 pc, and the mass obtained is $0.27 \pm 0.07$ $M_{\odot}$,
which is much lower than one would expect for an M1V pair. 

\noindent
{\bf YSC 156.} An F3V star with small magnitude difference, the expected mass for this system would be about 3 $M_{\odot}$. 
We derive $6.8 \pm 4.2$ $M_{\odot}$, so the orbit we present is not particularly useful in terms of stellar astrophysics,
but nonetheless should provide reasonable ephemerides for the coming few years.

\noindent
{\bf WSI 65 = 66 Oph.} This is a well-known Be star, and it is interesting to note that the orbit appears to be somewhat eccentric.
We obtain total mass of $10.4 \pm 3.8$ $M_{\odot}$. Given the spectral type of B2Ve and a $\Delta m$ of $\sim$3, we would 
expect that the secondary is an early A star; this combination would have a mass of $\sim$14 $M_{\odot}$, consistent with 
what we derive from the astrometry. The {\it Hipparcos} revised parallax is $5.01 \pm 0.26$ mas, so that the semi-major axis is
$34.9 \pm 3.8$ AU. We note that Draper \ea (2014) show that this star's $V$-band polarization has been declining over the last thirty
years; the last time of periastron passage was in 2003.1 according to our orbital elements, so this has been occurring when the
secondary is closest to the primary, suggestive of an effect by the secondary on the disk thought to surround the primary.

\subsection{Spectroscopic-visual Orbits}

The number of velocities from our spectroscopic observations of
HD 22451 = YSC~127 and HD 185501 = YSC~135, and their orbital phase coverage,
are sufficient to obtain spectroscopic orbital elements. We first determined
preliminary elements of the components of the two systems with the
program BISP (Wolfe \ea 1967), which uses the Wilsing-Russell Fourier
analysis method (Wilsing 1893, Russell 1902). We refined those elements with
the differential corrections program SB1 (Barker \ea 1967). We compared
the variances of the solutions for the primary and secondary of
YSC~127. The solution for the primary velocities has the smallest
variance, so we assigned unit weights to those velocities.
We then set the weights of the secondary velocities to be the inverse
of the ratio of the variances from the primary and secondary solutions.
Thus, for YSC~127 we assigned weights of 1.0 to the primary velocities
and 0.6 to those of the secondary except for the KPNO observation of
JD 2456868, which had very blended components measured as a single
velocity, and so was given a weight of zero. The two components of
YSC~135 have lines of very similar width and depth resulting in
nearly identical variances for their solutions, and so all velocities
for that system were given unit weights.
To obtain a simultaneous solution of the components of the two systems,
we adopted the above weights and used the program SB2, which is a
slightly modified version of SB1.
The spectroscopic orbital elements for both systems are listed in
Table~\ref{tbl-orbelem}. The resulting spectroscopic orbit for YSC~127 is
shown in Figure~\ref{22451sborb} and that for YSC~135 is displayed in
Figure~\ref{185501sborb}.

We next used our spectroscopic orbital elements as the starting point for combined spectroscopic-visual orbits
of the two systems, where the relative astrometry is entirely from our own speckle program at WIYN, Gemini, and the LDT. 
In this case the same method as in Muterspaugh \ea (2010) was used to 
do the simultaneous fitting. Results of these calculations are also shown in Table~\ref{tbl-orbelem}, including
the individual masses of the stars and the independently-determined distance to the system, which can be obtained 
when both astrometric and double-lined spectroscopic data exist. Figures 9 and 10 show the
visual orbits for YSC 127 and YSC 135, which have periods of
2401 days and 433.9 days, respectively.
The combined fit shows that for YSC 135, the standard deviation of separation residuals is 1.9 mas, comparable to 
the value estimated for the speckle observations in Section 4.1. However, in the case of YSC 127, the standard deviation in
$\rho$ is much higher, 5.4 mas. This would appear to be due primarily to the point included at a position angle of 325$^{\circ}$ in
2011; these were extremely challenging measurements made with the WIYN telescope at a separation of approximately
one-quarter of the diffraction limit and show that the separation
was probably overestimated in this case. 
Otherwise, the residuals for this system appear to be on the same level as for
YSC 135.

The distance obtained from the orbit calculation for YSC 127 is $114.6 \pm 6.3$ pc, whereas the value implied 
by the {\it Gaia} DR2 parallax is $126.2 \pm 0.7$ pc. Thus the 1-$\sigma$ error bars of the two measurements 
do not overlap at this stage. We attempted to compute an orbit without the 2011 data points and this resulted 
in a semi-major axis that was lower than the value shown in Table 9 by 6\%. This translates into a distance that 
is 6\% larger than our calculated value, or 121.9 pc. This would reconcile the distances given our level
of uncertainty, but it will nonetheless be important to see if the {\it Gaia} value is revised downward at all in future data releases, 
or, if further astrometric data can be obtained in the coming years, a subsequent orbit calculation might confirm an 
increased distance from what we have obtained. In any case, 
the masses of 1.34 $M_{\sun}$ and 1.24 $M_{\sun}$ implied by our orbit agree well with those expected for an
F6V+F7V pair, as does the magnitude difference and absolute system magnitude.
According to the Geneva-Copenhagen catalog (Nordstr\"om \ea 2004), the system has near solar metallicity.

For YSC 135, a G5 pair with [Fe/H]= $-$0.28, the distances agree well between the calculation here and DR2; the former is
$33.09 \pm 0.74$ pc while the latter is $32.58 \pm 0.03$ pc. The individual masses are 0.90 $M_{\sun}$ and
0.88 $M_{\sun}$ and are certainly roughly in line with stars of this
spectral type, if slightly low. This may be due to a combination of two factors. As shown in Horch \ea (2019), 
one would expect the value to be low compared to an equivalent binary of solar metallicity by 5-10\% at [Fe/H] = $-$0.3. On the
other hand, the $B-V$ color of the pair is shown to be 0.77 in SIMBAD, which is redder than expected for a G5 pair.

Torres et al.\ (2010) compiled a list of eclipsing binary
systems with masses and radii known to 3\% or better.
They also included 23 systems with accurate interferometric
and spectroscopic orbits that resulted in component
masses determined to 3\% or better. Our results for YSC 127
and YSC 135 indicate that the stars of these two systems
can now be added to the latter list.

\section{Conclusions}

We have presented 370 measures of binary star systems and results on 
76 further systems that show no evidence 
of a companion using speckle imaging at Lowell Observatory's Discovery Channel Telescope. Individual measures 
have relative astrometry that is precise to a level below 2 mas in separation and 0.6 degrees in position angle. 
Magnitude differences of this data set appear to be precise to approximately 0.15 magnitudes. There appear to be no
measurable offsets in our measures when comparing to other well-calibrated data sets.

While our survey of Be stars did not yield any previously unknown companions, we were able to calculate a first visual orbit for a 
Be star in one case, namely WSI 65 = 66 Oph.
The data presented here also include 15 previously unknown components from our observations of {\it Hipparcos} suspected binaries
and other stars, five of which were found in systems already known to be binary,
hence they are now revealed to be trinary systems. Of the remaining 10 systems, judging from magnitude differences 
and separations, it would appear likely that the majority are companions that are gravitationally bound to the primary star.
Combining the data here with previous data in the literature, we calculate six
visual orbits and two visual+spectroscopic orbits. In the latter case, individual masses are obtained to the 2\% level and 
the distance derived is consistent with DR2 in one case (YSC 135)
but discrepant at the 2-$\sigma$ level in the other (YSC 127), though this may be due to an overestimate of the separation
in the case of two observations taken at extremely small separation in 2011.

\acknowledgments
We thank the excellent team of telescope operators who facilitated these observations at the LDT: Heidi Larson, Teznie Pugh, 
Jason Sanborn, and Ana Hayslip. We are also grateful to Karen Bjorkman at the University of Toledo for her advice in 
the selection of the Be stars and aid in securing some of the telescope
time used here. We gratefully acknowledge support from the National Science Foundation, specifically, grants AST-1517824, AST-1616698, and AST-1909560 at Southern Connecticut State University, 
AST-1517413 at Georgia State University, and 
AST-1616084 at Lowell Observatory. 
JGW is supported by a
grant from the Templeton Foundation. The opinions expressed here are
those of the authors and do not necessarily reflect the views of the
John Templeton Foundation.
We made use of
the Washington Double Star Catalog maintained at the U.S. Naval Observatory, the
SIMBAD database, operated at CDS, Strasbourg, France, the 9th Catalog of
Spectroscopic Orbits of Binary Stars, maintained by D. Pourbaix.
We also used data from the European Space Agency (ESA) mission
{\it Gaia} (\url{https://www.cosmos.esa.int/gaia}), processed by the {\it Gaia}
Data Processing and Analysis Consortium (DPAC,
\url{https://www.cosmos.esa.int/web/gaia/dpac/consortium}). Funding for the DPAC
has been provided by national institutions, in particular the institutions
participating in the {\it Gaia} Multilateral Agreement.

% References -----------------------------------------------------------

\clearpage

% Tables -----------------------------

\begin{centering}

\begin{deluxetable}{lclrl}
\tabletypesize{\scriptsize}
\tablewidth{0pt}
\tablenum{1}
\tablecaption{Orbits Used in the DSSI Scale Determinations at the LDT}
\tablehead{
\colhead{Run} &
\colhead{WDS} &
\colhead{Discoverer} &
\colhead{HIP} &
\colhead{Orbit Reference} \\
&& \colhead{Designation} &&
}
\startdata

2015 March & $15232+3017$ & STF 1937AB & 75312 & Muterspaugh \ea 2010 \\
 & $15278+2906$ & JEF 1 & 75695 & Muterspaugh \ea 2010 \\

2015 June & $19490+1909$ & AGC 11AB & 97496 & Muterspaugh \ea 2010 \\
 
2015 November   & $04136+0743$ & A 1938 & 19719 &  Muterspaugh \ea 2010 \\
  & $21145+1000$ & STT 535AB & 104858 & Muterspaugh \ea 2008 \\
  & $22409+1433$ & HO 296AB & 111974 & Muterspaugh \ea 2010 \\

2016 January & $13100+1732$ & STF 1728AB & 64241 & Muterspaugh \ea 2015 \\

2016 February &  $04590-1623$ & BU 314AB & 23166 & this paper \\
 
2017 May & $13100+1732$ & STF 1728AB & 64241 & Muterspaugh \ea 2015 \\
 & $15232+3017$ & STF 1937AB & 75312 & Muterspaugh \ea 2010 \\
 & $15278+2906$ & JEF 1 & 75695 & Muterspaugh \ea 2010 \\
 & $17080+3556$ & HU 1176AB & 83838 & Muterspaugh \ea 2010 \\
\enddata
 
 \end{deluxetable}

\begin{deluxetable}{lcllc}
\tabletypesize{\footnotesize}
\tablewidth{0pt}
\tablecolumns{5}
\tablenum{2}
\tablecaption{Telescope and Instrument Combinations for Spectroscopic Observations\tablenotemark{a} \label{tbl-instcomb}}
\tablehead{\colhead{Date} & \colhead{Helio. Julian Date} & \colhead{Telescope} &
\colhead{Instrument,} & \colhead{Resolution} \\
\colhead{d/m/yr} & \colhead{HJD$-2400000$} & \colhead{} & \colhead{grating,CCD} &
\colhead{$\lambda/\Delta \lambda$ (2 pixel)}}
\startdata
09 10 2012 & 56209 & KPNO coud\'e feed & coud\'e spec., echelle, T2KB & 72000 \\
30 11 2013 & 56261 & KPNO 4 m & echelle spec., 58-63,T2KA & 41000 \\
20 04 2013 & 56402 & KPNO coud\'e feed & coud\'e spec., A, STA3 & 26600 \\
22 05 2013 & 56434 & KPNO coud\'e feed & coud\'e spec., A, STA3 & 26600 \\
26 10 2013 & 56591 & KPNO coud\'e feed & coud\'e spec., echelle ,T2KB & 72000 \\
07 01 2014 & 56664 & KPNO coud\'e feed & coud\'e spec., echelle ,T2KB & 72000 \\
24 04 2014 & 56771 & KPNO coud\'e feed & coud\'e spec., echelle ,T2KB & 72000 \\
30 07 2014 & 56868 & KPNO coud\'e feed & coud\'e spec., echelle ,T2KB & 72000 \\
03 04 2015\tablenotemark{b}& 57115 & TSU AST 2 m & spec., echelle, Fairchild486 & 25000 \\
27 05 2016 & 57535 & KPNO WIYN 3.5 m & Hydra, echelle, STA2 & 16500 \\
02 01 2018 & 58120 & KPNO WIYN 3.5 m & Hydra, echelle, STA2 & 16500 \\
24 10 2018 & 58415 & KPNO WIYN 3.5 m & Hydra, echelle, STA2 & 16500 \\
\enddata
\tablenotetext{a}{See Tables~\ref{tbl-22451obs} and \ref{tbl-185501obs}
for a complete list of dates, velocities, and sources.}
\tablenotetext{b}{First observation of AST series.}

\end{deluxetable}

\begin{deluxetable}{cccrccrcl}
\tabletypesize{\scriptsize}
\tablenum{3}
\tablewidth{0pt}
\tablecaption{Radial Velocity Observations of HD 22451 = YSC 127 \label{tbl-22451obs}}
\tablehead{ \colhead{Helio. Julian Date} & \colhead {Phase} &
\colhead{$V_{A}$} & \colhead{$(O-C)_{A}$} & \colhead{Weight$_{A}$} &
\colhead{$V_{B}$} & \colhead{$(O-C)_{B}$} & \colhead{Weight$_{B}$} &
\colhead{Source\tablenotemark{a}} \\
\colhead{(HJD $-$ 2400000)} & \colhead {} & \colhead{(km~s$^{-1}$)} &
\colhead{(km~s$^{-1}$)}  & \colhead{} & \colhead{(km~s$^{-1}$)} &
\colhead{(km~s$^{-1}$)} & \colhead{} & \colhead{}
}
\startdata
 56209.8155 & 0.117 & $-$30.9 & $-$0.2 & 1.0 &  $-$6.9 & $-$0.1 & 0.6 & KPNO \\
 56261.8133 & 0.139 & $-$29.4 &    0.1 & 1.0 &  $-$8.3 & $-$0.2 & 0.6 & KPNO \\
 56591.8312 & 0.280 & $-$23.2 &    0.3 & 1.0 & $-$14.0 &    0.6 & 0.6 & KPNO \\
 56664.6869 & 0.311 & $-$22.5 &    0.0 & 1.0 & $-$15.0 &    0.7 & 0.6 & KPNO \\
 56868.9608 & 0.398 & $-$19.7 &    0.3 & 0.0 & $-$19.7 & $-$1.3 & 0.0 & KPNO \\
 57167.9708 & 0.525 & $-$16.3 &    0.6 & 1.0 & $-$22.5 & $-$0.8 & 0.6 & Fair \\
 57401.8338 & 0.625 & $-$14.4 &    0.3 & 1.0 & $-$24.2 &    0.0 & 0.6 & Fair \\
 57434.7350 & 0.639 & $-$14.4 & $-$0.1 & 1.0 & $-$24.6 & $-$0.1 & 0.6 & Fair \\
 57464.7278 & 0.651 & $-$14.7 & $-$0.6 & 1.0 & $-$25.7 & $-$0.9 & 0.6 & Fair \\
 57470.7050 & 0.654 & $-$14.4 & $-$0.4 & 1.0 & $-$25.7 & $-$0.8 & 0.6 & Fair \\
\enddata
\tablecomments{$^a$KPNO = Kitt Peak National Observatory, Fair = Fairborn
Observatory. \\ (Table 3 is published in its entirety in the machine-readable format.
      A portion is shown here for guidance regarding its form and content.)}

\end{deluxetable}

\begin{deluxetable}{cccrccrcl}
\tabletypesize{\scriptsize}
\tablenum{4}
\tablewidth{0pt}
\tablecaption{Radial Velocity Observations of HD 185501 = YSC 135 \label{tbl-185501obs}}
\tablehead{ \colhead{Helio. Julian Date} & \colhead {Phase} &
\colhead{$V_{A}$} & \colhead{$(O-C)_{A}$} & \colhead{Weight$_A$} &
\colhead{$V_{B}$} & \colhead{$(O-C)_{B}$} & \colhead{Weight$_B$} &
\colhead{Source\tablenotemark{a}} \\
\colhead{(HJD $-$ 2400000)} & \colhead {} & \colhead{(km~s$^{-1}$)} &
\colhead{(km~s$^{-1}$)} & \colhead{} &\colhead{(km~s$^{-1}$)} &
\colhead{(km~s$^{-1}$)} & \colhead{} & \colhead{}
}
\startdata
 56209.6237 & 0.878 & $-$16.3 &    0.4 & 1.0 & $-$47.2 &    0.4 & 1.0 & KPNO \\
 56402.9917 & 0.323 & $-$43.7 &    0.1 & 1.0 & $-$20.3 & $-$0.5 & 1.0 & KPNO \\
 56434.9579 & 0.397 & $-$40.5 & $-$0.4 & 1.0 & $-$23.6 &    0.0 & 1.0 & KPNO \\
 56591.6410 & 0.757 & $-$18.3 & $-$0.2 & 1.0 & $-$46.1 &    0.0 & 1.0 & KPNO \\
 56771.9861 & 0.172 & $-$46.3 &    0.2 & 1.0 & $-$17.0 &    0.1 & 1.0 & KPNO \\
 56868.7441 & 0.395 & $-$40.1 &    0.1 & 1.0 & $-$23.5 &    0.0 & 1.0 & KPNO \\
 57115.9458 & 0.964 & $-$23.5 &    0.4 & 1.0 & $-$40.1 &    0.1 & 1.0 & Fair \\
 57160.9458 & 0.067 & $-$38.9 & $-$0.1 & 1.0 & $-$24.9 &    0.0 & 1.0 & Fair \\
 57161.8844 & 0.070 & $-$39.2 & $-$0.1 & 1.0 & $-$24.7 &    0.0 & 1.0 & Fair \\
 57184.7764 & 0.122 & $-$44.1 &    0.1 & 1.0 & $-$19.7 & $-$0.3 & 1.0 & Fair \\
\enddata
\tablecomments{$^a$KPNO = Kitt Peak National Observatory, Fair = Fairborn
Observatory. \\ (Table 4 is published in its entirety in the machine-readable format.
      A portion is shown here for guidance regarding its form and content.)}
\end{deluxetable}

\clearpage

\pagestyle{empty}

\begin{deluxetable}{lllrlrrrrl}
\tabletypesize{\scriptsize}
\tablewidth{0pt}
\tablenum{5}
\tablecaption{Binary star speckle measures}
\tablehead{ 
\colhead{WDS} & 
\colhead{HR,ADS} & 
\colhead{Discoverer} & HIP & 
\colhead{Date} & 
\colhead{$\theta$} & \colhead{$\rho$} & 
\colhead{$\Delta m$} & 
\colhead{$\lambda$} & 
\colhead{$\Delta \lambda$} \\
\colhead{($\alpha$,$\delta$ J2000.0)} & 
\colhead{DM,etc.} & \colhead{Designation} && 
\colhead{(2000+)} & 
\colhead{($^{\circ}$)} & 
\colhead{(${\prime \prime}$)} 
 & \colhead{(mag)}
 & \colhead{(nm)} & 
\colhead{(nm)} 
} 
\startdata 

$01095+4715$ & ADS 940 & STT 515AB  &    5434 & 
   15.8338 & 116.6 & 0.5225 & 1.40 & 692 & 40 \\
&&&&
   15.8338 & 116.2 & 0.5209 & 1.46 & 880 & 50 \\

$02177+4235$ & HD 14064 & LSC 20 &   10695 & 
   15.8368 & 125.8 & 0.0535 & 2.15 & 692 & 40\tablenotemark{a} \\
&&&&
   15.8368 & 306.8 & 0.0572 & 1.90 & 880 & 50 \\

$02225-2349$ & HR 695 & LAF 27 &   11072 & 
   15.8368 &  94.1 & 0.2716 & 4.12 & 692 & 40 \\
&&&&
   15.8368 &  91.4 & 0.2716 & 3.73 & 880 & 50 \\

$02512+6023$ & ADS 2165 & BU 1316AB &   13308 & 
   15.8342 & 299.3 & 0.3198 & 0.76 & 692 & 40 \\
&&&&
   15.8342 & 298.8 & 0.3201 & 0.76 & 880 & 50 \\

$03117+3115$ & BD+30 500 & LSC 23 &   14840 & 
   15.8342 & 256.5 & 0.0771 & 2.65 & 692 & 40 \\
&&&&
   15.8342 & 252.5 & 0.0787 & 2.02 & 880 & 50 \\

\enddata 

\tablenotetext{a}{Quadrant ambiguous.} 
%\tablenotetext{b}{Quadrant inconsistent with previous measures in the 4th Interferometric Catalog (Hartkopf \ea 2001b).}
%\tablenotetext{c}{Photometry for this observation appears as an upper limit because
%the observation may be affected by speckle decorrelation as discussed in Section 4.3.}

%\tablenotetext{c}{A magnitude difference for this single-lined spectroscopic binary does not
%appear for the reasons discussed in the text for small-separation systems.}

\tablecomments{Table 5 is published in its entirety in the machine-readable format.
      A portion is shown here for guidance regarding its form and content.}

\end{deluxetable} 

%\clearpage

\begin{deluxetable}{clrcrrrrl}
%\tabletypesize{\scriptsize}
\tabletypesize{\tiny}
\tablewidth{0pt}
\tablenum{6}
\tablecaption{Ephemeris Positions and Residuals Used in the Astrometric Accuracy Study}
\tablehead{
\colhead{WDS} &
\colhead{Discoverer} &
\colhead{HIP} &
\colhead{Date} &
\colhead{$\theta_{eph}$} &
\colhead{$\rho_{eph}$} &
\colhead{${\Delta \bar{\theta}}$} &
\colhead{${\Delta \bar{\rho}}$} &
\colhead{WDS Orbit Grade} \\
& \colhead{Designation} &&
\colhead{($2000+$)} &
\colhead{($^{\circ}$)} &
\colhead{($^{\prime \prime}$)} &
\colhead{($^{\circ}$)} &
\colhead{(mas)} & \colhead{and Reference}
}
\startdata
%$02396-1152$ & FIN 312 &  12390 & 17.9477 &
 % $200.9 \pm 0.4$ & $0.0855 \pm 0.0004$ & $+2.4$ & $-2.5$ & 1, Docobo \& Andrade (2013) \\
% Orbit: Doc2013d
% Docobo, J. A. and Andrade, M., MNRAS 428, 321, 2013

$07480+6018$ & HU 1247 & 38052 & 15.1848 & 
  $3.3 \pm 1.5$ & $0.1654 \pm 0.0018$ & $-$2.2 & $-$1.0 & 2, Hartkopf \ea (1996) \\
 
 $07518-1354$ & BU 101 & 38382 & 15.1849 &
   $294.5 \pm 0.1$ & $0.5554 \pm 0.0027$ & $+$0.1 & $+$3.8 & 1, Tokovinin (2012) \\
 
$07528-0526$ & FIN 325 & 38474 & 15.1849 &
  $182.0 \pm 0.3$ & $0.3506 \pm 0.0036$ & $-$0.3 & $+$2.4 & 2, Hartkopf \ea (1996) \\

$09036+4709$ & A 1585 & 44471 & 15.1825 & 
  $286.5 \pm 0.2$ & $0.2830 \pm 0.0004$ & $+$0.1 & $+$0.3 & 2, Muterspaugh \ea (2010) \\
 
%$09123+1500$ & FIN 347 & 45170 & 15.1852 &
%315.5 & 0.0686 & & & 1, Msn2012a \\

%$09474+1134$ & MCA 34 & 48029 & 15.1855 &
% & & & & 2, \\
 
$10083+3136$ & KUI 48 & 49658 & 15.1827 &
  $169.8 \pm 0.2$ & $0.2171 \pm 0.0009$ & $+$0.5 & $-$3.1 & 2, Hartkopf \ea (1996) \\
 
 $12060+6842$ & STF  3123 & 59017 & 15.1857 &
   $197.1 \pm 3.7$ & $0.2991 \pm 0.0072$ & $-$0.8 & $-$10.9 & 2, Hartkopf \ea (1996)\\
  
% $12199-0040$ & MCA 37 & 60129 &15.1829 &
% & & & & 2, \\
  
 $12417-0127$ & STF 1670AB & 61941 & 15.1829 &
   $5.6 \pm 0.6$ & $2.2955 \pm 0.0055$ & $-$0.1 & $-$23.8 & 2, Scardia \ea (2007) \\
  
 $12417-0127$ & STF 1670AB & 61941 & 16.0720 &
   $3.7 \pm 0.5$ & $2.4245 \pm 0.0058$ & $-$0.3 & $-$10.7 & 2, Scardia \ea (2007) \\

%$13175-0041$ & FIN 350 & 64838 & 15.1829 &
% & & & & 2, \\
 
 $13198+4747$ & HU 644AB & 65026 & 15.1828 & 
   $86.1 \pm 0.4$ & $0.7996 \pm 0.0301$ & $+$0.1 & $+$36.5 & 2, Hartkopf and Mason (2015) \\
 
 $13203+1746$ & A 2166 & 65069 & 15.1829 &
   $6.8 \pm 5.3$ & $0.1313 \pm 0.0292$ & $-$1.3 & $+$4.5 & 2, Zasche and Uhlar (2010) \\

\enddata

%\tablenotetext{a}{Figures in parentheses did not meet the criteria explained in the text for comparison.}

\end{deluxetable}

\clearpage

\begin{deluxetable}{cccccccl}
\tabletypesize{\tiny}
\tablewidth{0pt}
\tablenum{7}
\tablecaption{5-$\sigma$ Detection Limits for High-Quality Nondetections}
\tablehead{
\colhead{($\alpha$,$\delta$ J2000.0)} & \colhead{\it Hipparcos} &
\colhead{Date} &
\multicolumn{2}{r}{5-$\sigma$ Det. Lim., 692 nm} &
\multicolumn{2}{r}{5-$\sigma$ Det. Lim., 880 nm} & 
\colhead{List, Notes} \\
\colhead{(WDS format)} & Number &
\colhead{(Bess. Yr.)} &
\colhead{0.2$^{\prime \prime}$} &
\colhead{1.0$^{\prime \prime}$} &
\colhead{0.2$^{\prime \prime}$} &
\colhead{1.0$^{\prime \prime}$} 
}
\startdata
$00064+6412$ &    531 & 15.8338 & 3.80  &  5.96  &  3.76  &  6.23 & Be Star (10 Cas) \\
$00447+4817$ &   3504 & 15.8338 & 3.77  &  5.59  &  3.83  &  6.01 & Be Star ($o$ Cas) \\
$00567+6043$ &   4427 & 15.8338 & 3.66  &  5.66  &  4.38  &  6.12 & Be Star ($\gamma$ Cas) \\
$01557+5916$ &   8980 & 15.8340 & 4.39  &  6.14  &  4.20  &  6.38 & Be Star (V777 Cas) \\
$03365+4812$ &  16826 & 15.1846 & 4.23  &  6.34  &  4.27  &  5.99 & Be Star ($\psi$ Per) \\
$03365+4812$ &  16826 & 15.8342 & 4.35  &  7.00  &  3.64  &  7.06 & Be Star ($\psi$ Per) \\
$03423+1942$ &  17309 & 15.8342 & 4.79  &  7.21  &  4.38  &  7.49 & Be Star (13 Tau) \\
$03449+2407$ &  17499 & 15.8342 & 4.24  &  7.49  &  4.22  &  7.80 & Be Star (17 Tau) \\
$03463+2357$ &  17608 & 15.8343 & 4.33  &  7.22  &  4.12  &  7.65 & Be Star (23 Tau) \\
$03475+2406$ &  17702 & 15.8343 & 4.31  &  7.23  &  4.38  &  7.66 & Be Star ($\eta$ Tau) \\
\enddata
\tablecomments{Table 7 is published in its entirety in the machine-readable format.
      A portion is shown here for guidance regarding its form and content.}
\end{deluxetable}

\begin{deluxetable}{lrrrrrrrr}
%\tabletypesize{\scriptsize}
\tabletypesize{\small}
\tablewidth{0pt}
\tablenum{8}
\tablecaption{Visual Orbital Elements for Six Systems}
\tablehead{ 
\colhead{Name} &
\colhead{HIP} &
\colhead{$P$} &
\colhead{$a$} &
\colhead{$i$} &
\colhead{$\Omega$} &
\colhead{$T_{0}$} &
\colhead{$e$} &
\colhead{$\omega$} \\
&& \colhead{(yr)} &
\colhead{($^{\prime \prime}$)} &
\colhead{($^{\circ}$)} &
\colhead{($^{\circ}$)} &
\colhead{(BY)} &
\colhead{} &
\colhead{($^{\circ}$)}
} 
\startdata 

BU 314AB & 23166 & 59.78 & 0.4640 & 112.5 & 140.34 & 2037.95 & 0.845 & 357.0 \\
&& $\pm 0.64$ & $\pm 0.0048$ & $\pm 1.9$ & $\pm 0.82$ & $\pm 0.61$ & $\pm 0.018$ & $\pm 2.3$ \\

HDS 1199 & 41322 & 133.1 & 0.940 & 134.2 & 272 & 2007 & 0.110 & 71 \\
&& $\pm 6.7$ & $\pm 0.027$ & $\pm 2.5$ & $\pm 11$ & $\pm 34$ & $\pm 0.074$ & $\pm 29$ \\

COU 1258 & 48572 & 96 & 0.198 & 85.29 & 232 & 2096 & 0.237 & 259 \\
&& $\pm 13$ & $\pm 0.017$ & $\pm 0.87$ & $\pm 26$ & $\pm 39$ & $\pm 0.092$ & $\pm 44$ \\

HDS 1542AB & 52774 & 121.7 & 0.518 & 60.5 & 235.0 & 2100.6 & 0.412 & 250.4 \\
&& $\pm 8.2$ & $\pm 0.031$ & $\pm 2.7$ & $\pm 3.4$ & $\pm 8.2$ & $\pm 0.069$ & $\pm 6.8$ \\

YSC 156 & 82642 & 53 & 0.156 & 52 & 0 & 2038.1 & 0.16 & 175 \\
&& $\pm 11$ & $\pm 0.018$ & $\pm 14$ & $\pm 27$ & $\pm 3.7$ & $\pm 0.14$ & $\pm 33$ \\

WSI 65 & 88149 & 63.9 & 0.175 & 75.0 & 338.9 & 2067 & 0.37 & 117 \\
&& $\pm 5.0$ & $\pm 0.017$ & $\pm 1.0$ & $\pm 7.4$ & $\pm 12$ & $\pm 0.16$ & $\pm 51$ 

\enddata 

%\tablenotetext{a}{The spectroscopic elements are fixed to those of Latham \ea (2012).}

\end{deluxetable}

\begin{deluxetable}{lrrrr}
\tabletypesize{\footnotesize}
\tablenum{9}
\tablewidth{0pt}
\tablecaption{Orbital Elements of HD 22451 = YSC 127 and HD 185501 = YSC 135  \label{tbl-orbelem}
}
\tablehead{\colhead{Parameter} & \colhead{HD 22451} & \colhead{HD 22451}
& \colhead{HD 185501} & \colhead{HD 185501} \\
\colhead{} & \colhead{SB2} & \colhead{Joint VB+SB2 Fit} & \colhead{SB2} &
\colhead{Joint VB+SB2 Fit}
}
\startdata
$P$ (days)     & 2347.7 $\pm$ 6.3     & 2401.1 $\pm$ 3.2    & 434.57 $\pm$ 0.16  &  433.94 $\pm$ 0.15\\
$T$ (HJD)      & 2458283.0 $\pm$ 1.6      & 2458281.9 $\pm$ 1.5  & 2457132.14 $\pm$ 0.78  &  2457134.26 $\pm$ 0.62 \\
$e$                  & 0.5753 $\pm$ 0.0029  & 0.5804 $\pm$ 0.0030  & 0.2127 $\pm$ 0.0020 &  0.2155 $\pm$0.0023 \\
$a$ (mas)         & ...     & 42.0 $\pm$ 2.2        & ...        &  41.04 $\pm$ 0.88 \\
$i$ (deg)            & ...  & 111.29 $\pm$ 0.93     & ...        &  116.40 $\pm$ 0.93 \\
$\Omega$ (deg)  & ...       & 3.20 $\pm$ 0.82    & ...        &  337.59 $\pm$ 0.47 \\
$\omega{_A}$ (deg)  & 110.00 $\pm$ 0.33   & 109.51 $\pm$ 0.33    & 81.10 $\pm$ 0.69 & 82.92 $\pm$ 0.56 \\
$K_A$ (km s$^{-1}$)  & 12.001 $\pm$ 0.047   & ... & 15.422 $\pm$ 0.044 & ... \\
$K_B$ (km s$^{-1}$)  & 13.003 $\pm$ 0.053   &... & 15.801 $\pm$ 0.044  & ... \\
$\gamma$ (km s$^{-1}$) & $-$19.225 $\pm$ 0.027 & ... & -31.948 $\pm$ 0.025 & ...  \\
%$m_A\sin ^3i$ ($M_\odot$)  &        &  & $\pm$  &  \\
%$m_B\sin ^3i$ ($M_\odot$)  &        &  & $\pm$  &  \\
%$a_A\sin i$ ($10^6$\,km)   &        &  & $\pm$  &  \\
%$a_B\sin i$ ($10^6$\,km)   &        &  & $\pm$  &  \\
$M_{\rm tot}$ ($M_{\odot}$) & ... & 2.579 $\pm$ 0.054 & ... & 1.775 $\pm$ 0.023 \\
$M_{\rm A}$ ($M_{\odot}$) & ... & 1.342 $\pm$ 0.029 & ... & 0.898 $\pm$ 0.012 \\
$M_{\rm B}$ ($M_{\odot}$) & ... & 1.236 $\pm$ 0.026 & ... & 0.876 $\pm$ 0.012 \\
Distance (pc) & ... &  114.6 $\pm$ 6.3 & ... & 33.09 $\pm$ 0.74 \\
\enddata
\end{deluxetable}

\end{centering}

\clearpage

\begin{figure}[tb]
\plottwo{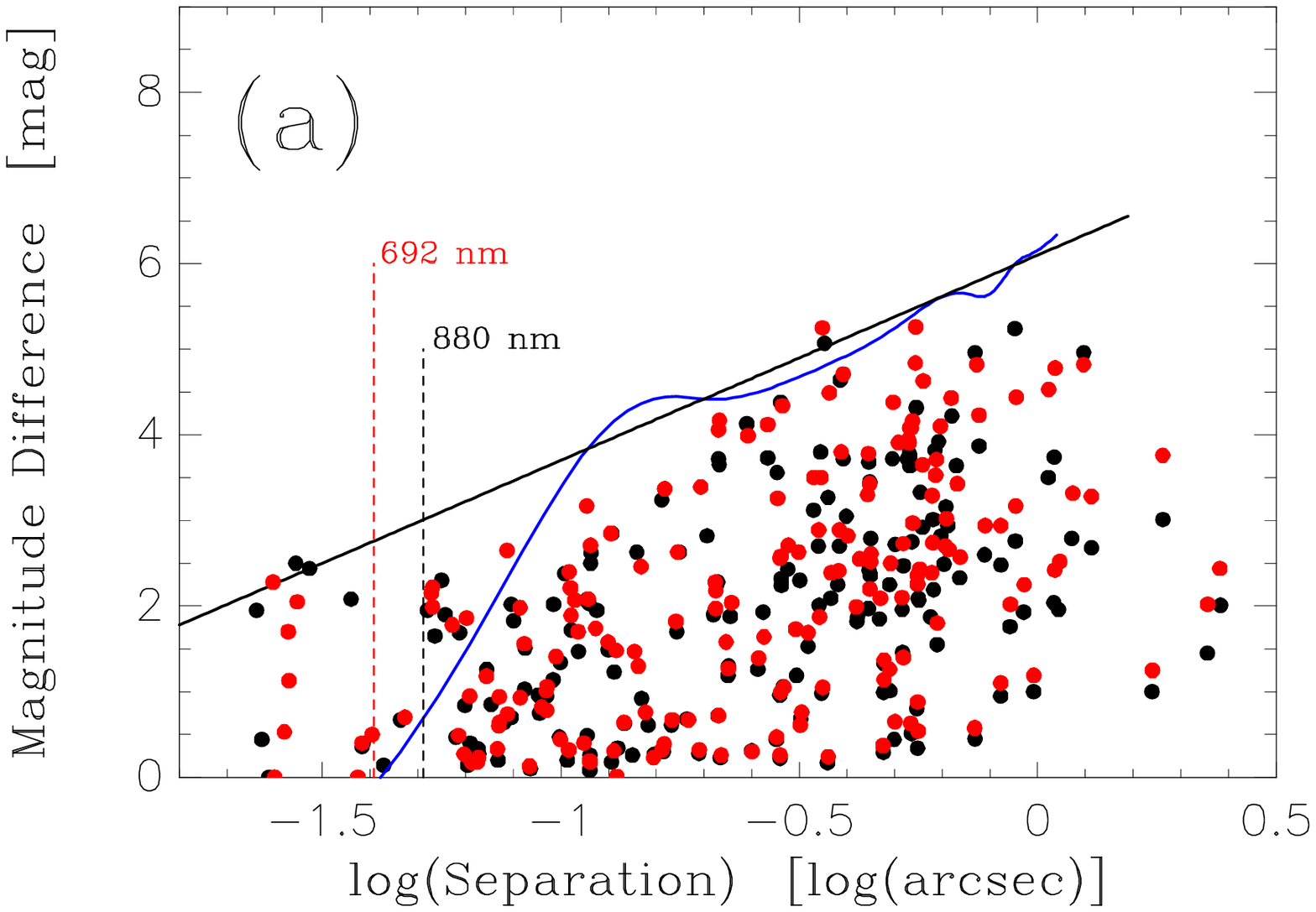}{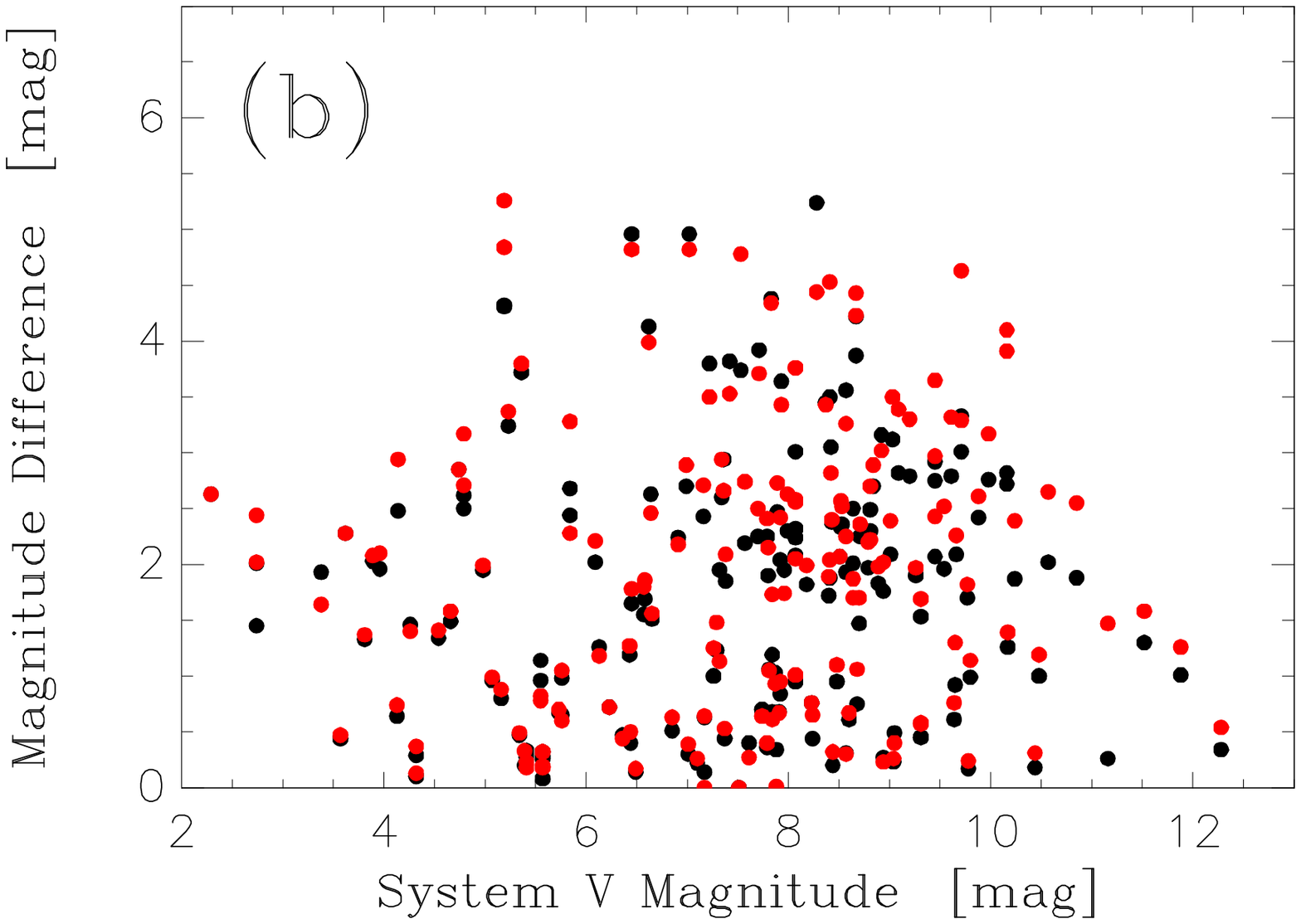}
\caption{(a) Magnitude difference measures in Table 5 are plotted as a function of the separation of the components. 
Black data points are results for the 880-nm filter, and red data points indicate results in the 692-nm filter. The blue 
line represents a typical detection limit curve obtained in good conditions when no companion is present, and the black 
line is chosen to approximately match the detection limit curve at larger separations, and then is simply extended into
the sub--diffraction-limited regime. The dashed vertical lines indicate the separation corresponding to the Rayleigh criterion 
for each wavelength.
(b) Magnitude difference measures plotted as a function of the system $V$ magnitude. Again, black points are used for
data at 880 nm, and red for data at 692 nm.}
\end{figure}

\begin{figure}[tb]
\plottwo{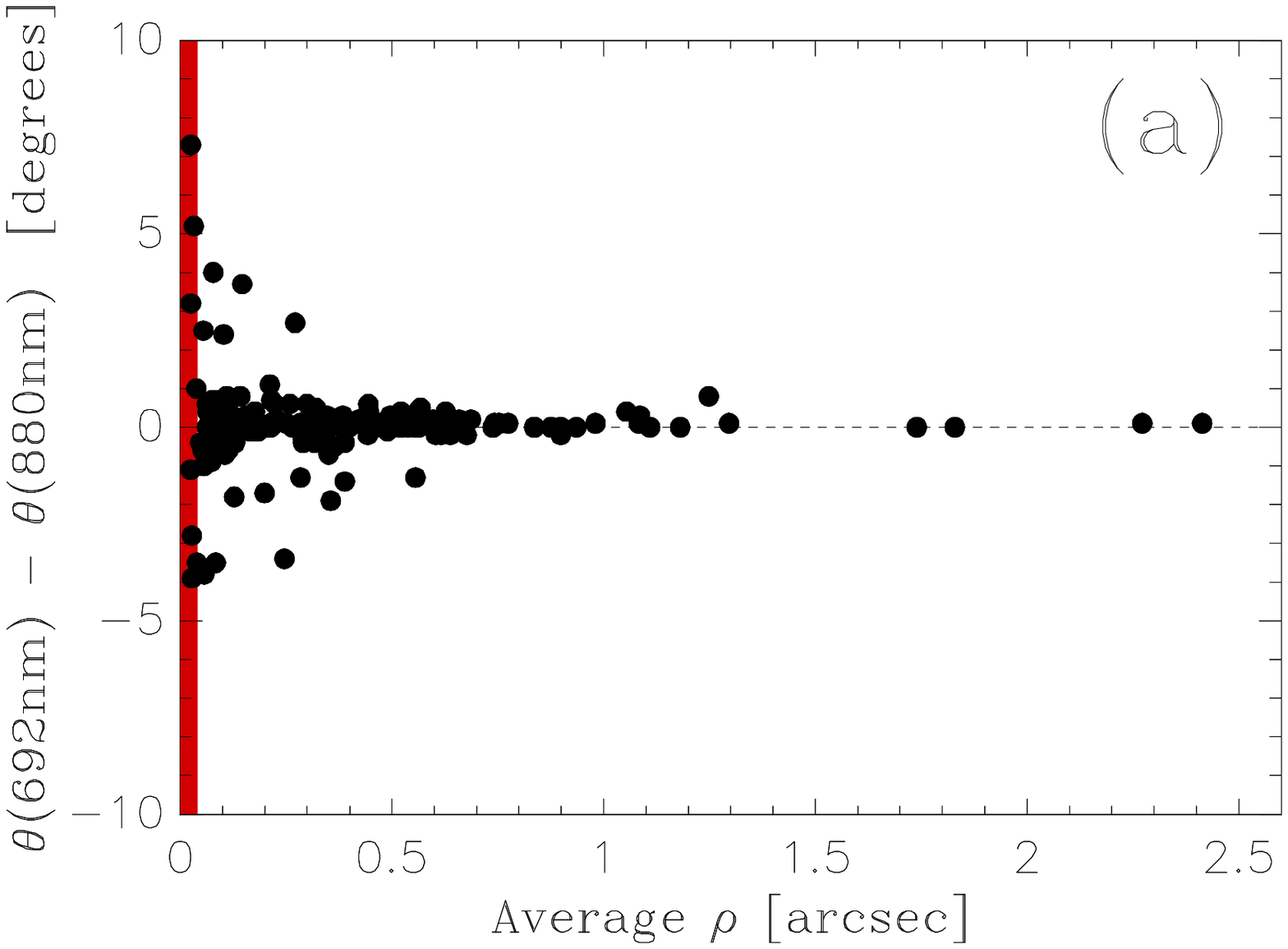}{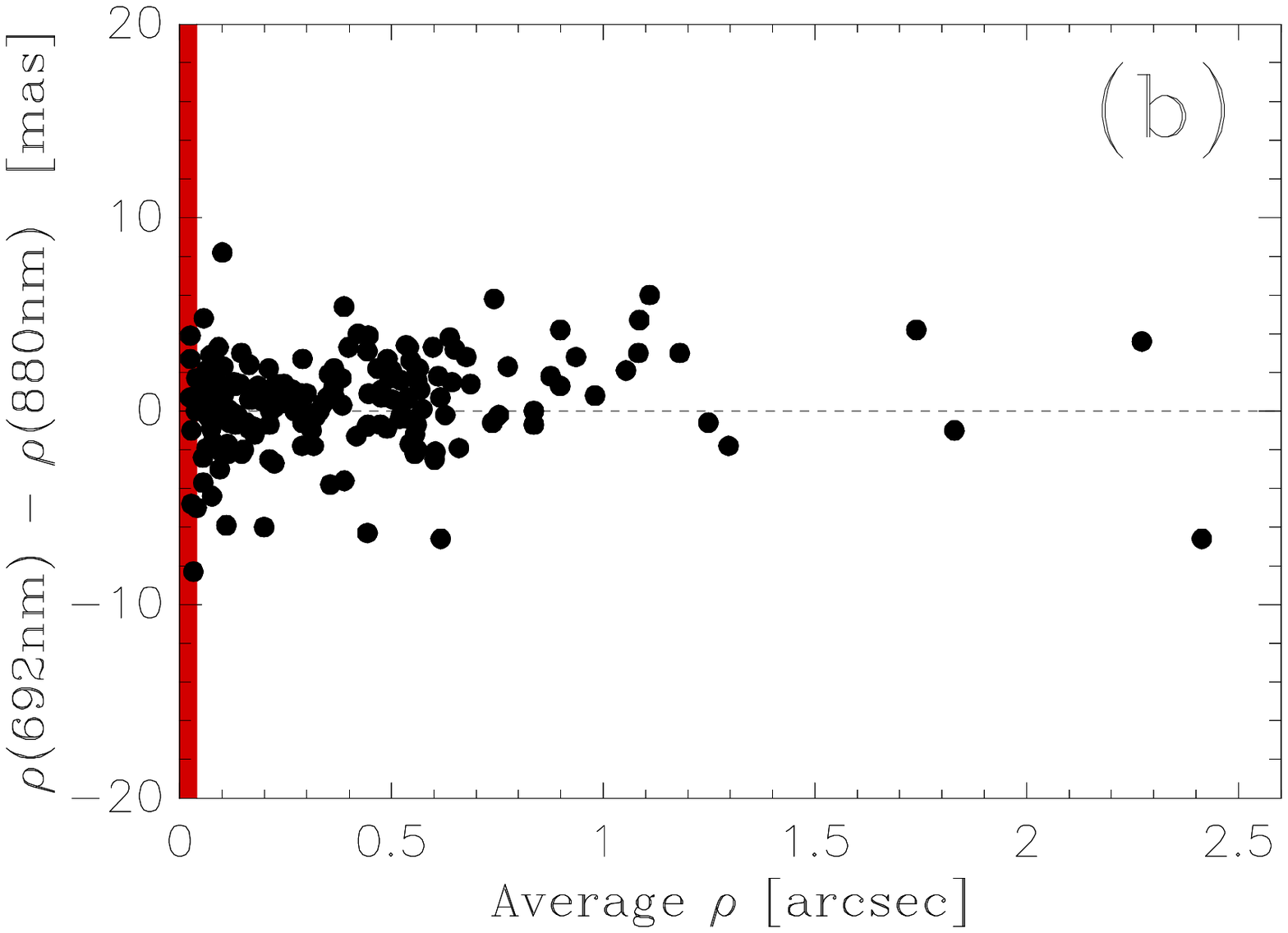}
\caption{(a) Differences in the position angle value obtained between the two channels of the instrument, as a function of the
separation of the system.
(b) Differences in the separation measure as a function of separation.}
\end{figure}

\clearpage

\begin{figure}[tb]
\plottwo{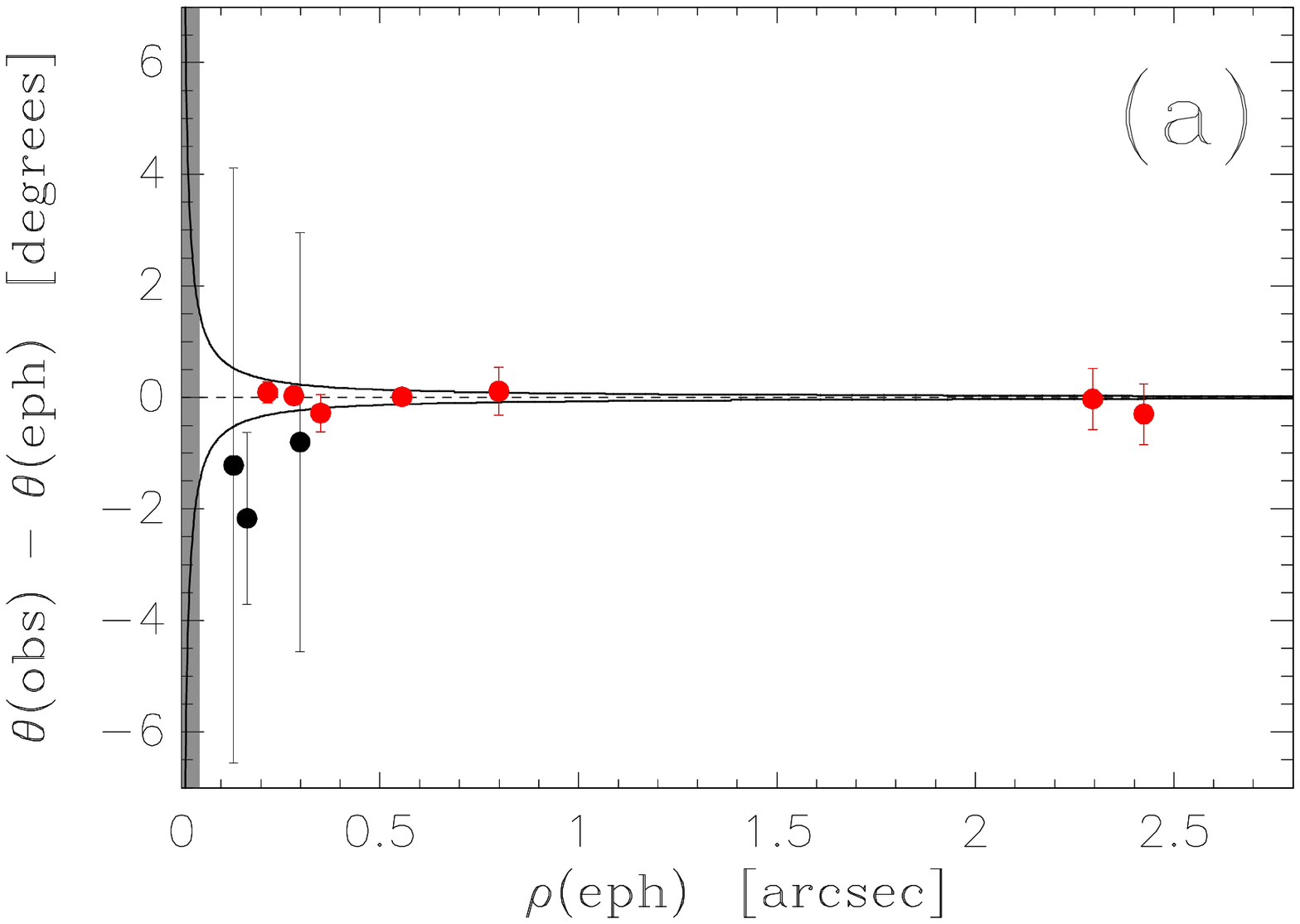}{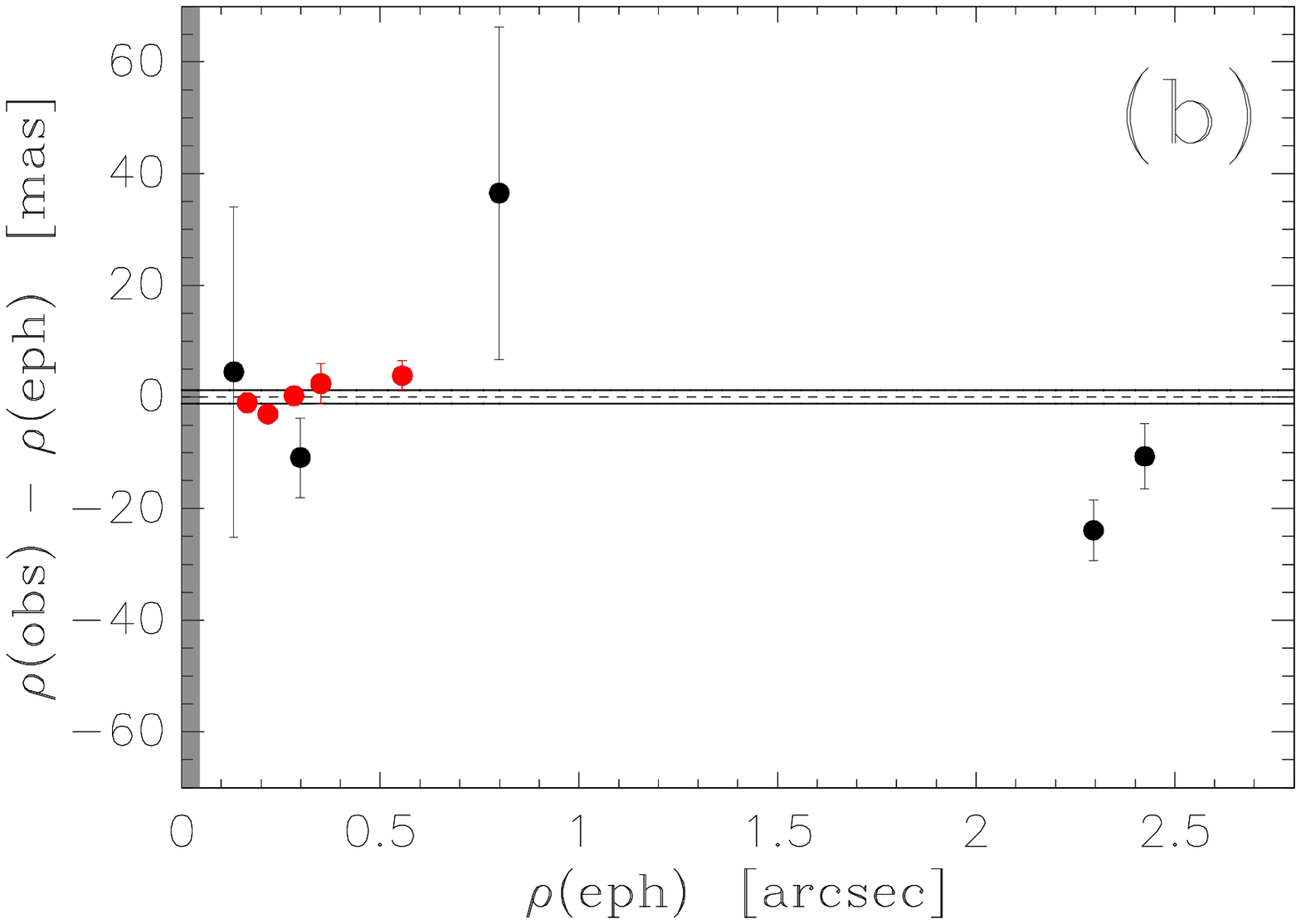}
\caption{ Observed minus ephemeris residuals obtained for measures of objects having Grade 1 or 2 orbits in the 
Sixth Orbit Catalog but not used in the scale determination. (a) Position angle residuals. 
The error bars shown are the uncertainties in the ephemeris 
positions as computed from the orbital elements, and the data points shown in red have uncertainties in the ephemeris 
position angle of less than 1$^{\circ}$. 
(b) Residuals in the separations for the same set of observations as a function of separation. In this case, objects with 
uncertainties in the ephemeris separations of less than 4 mas are shown in red. In both cases, the shaded region marks
the region below the diffraction limit of the telescope and the dashed line shows the zero line to guide the eye.
The curves drawn in (a) and the horizontal lines in (b) mark the level of internal precision described in the text.}
\end{figure}

\begin{figure}[tb]
\plottwo{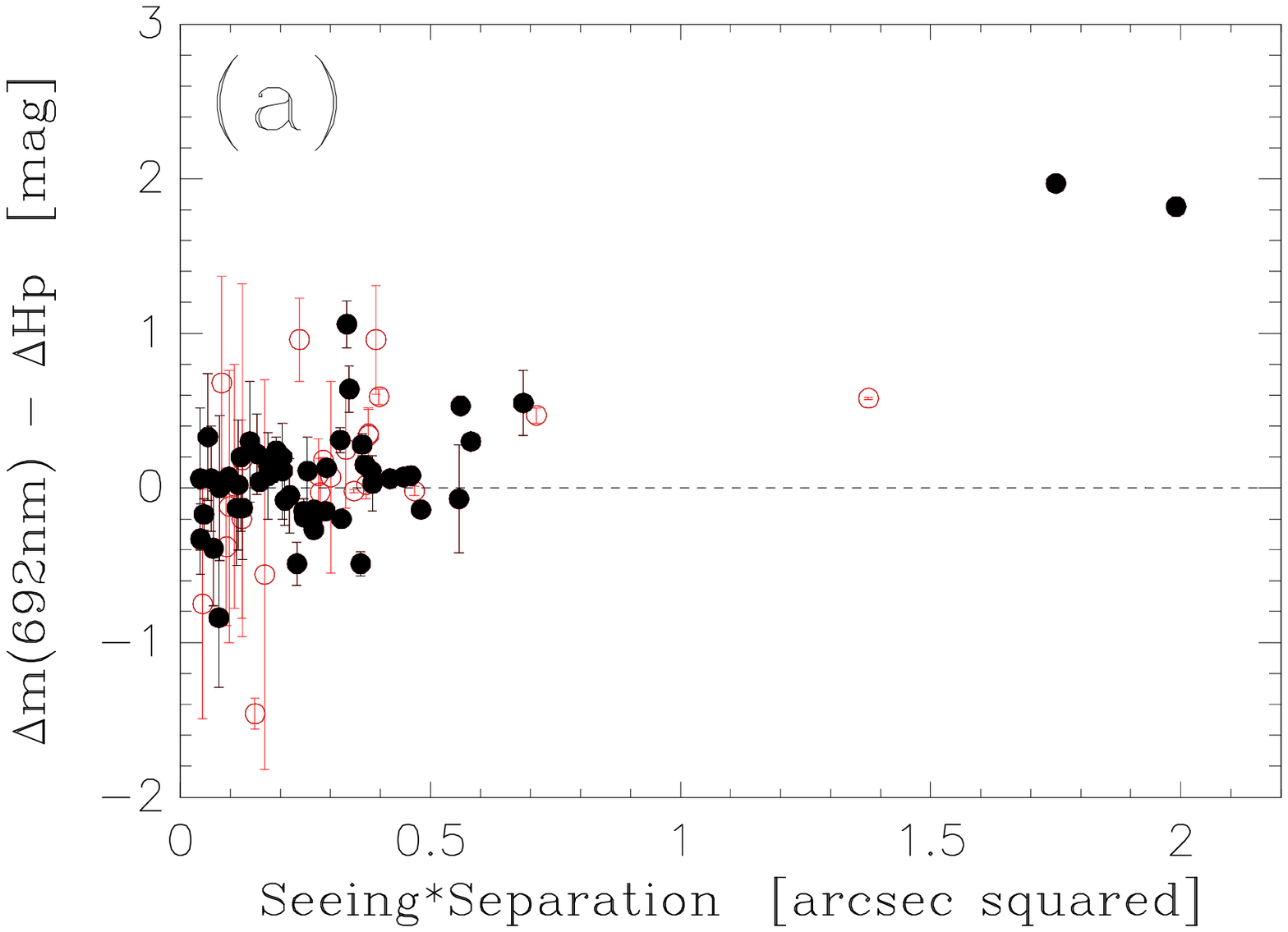}{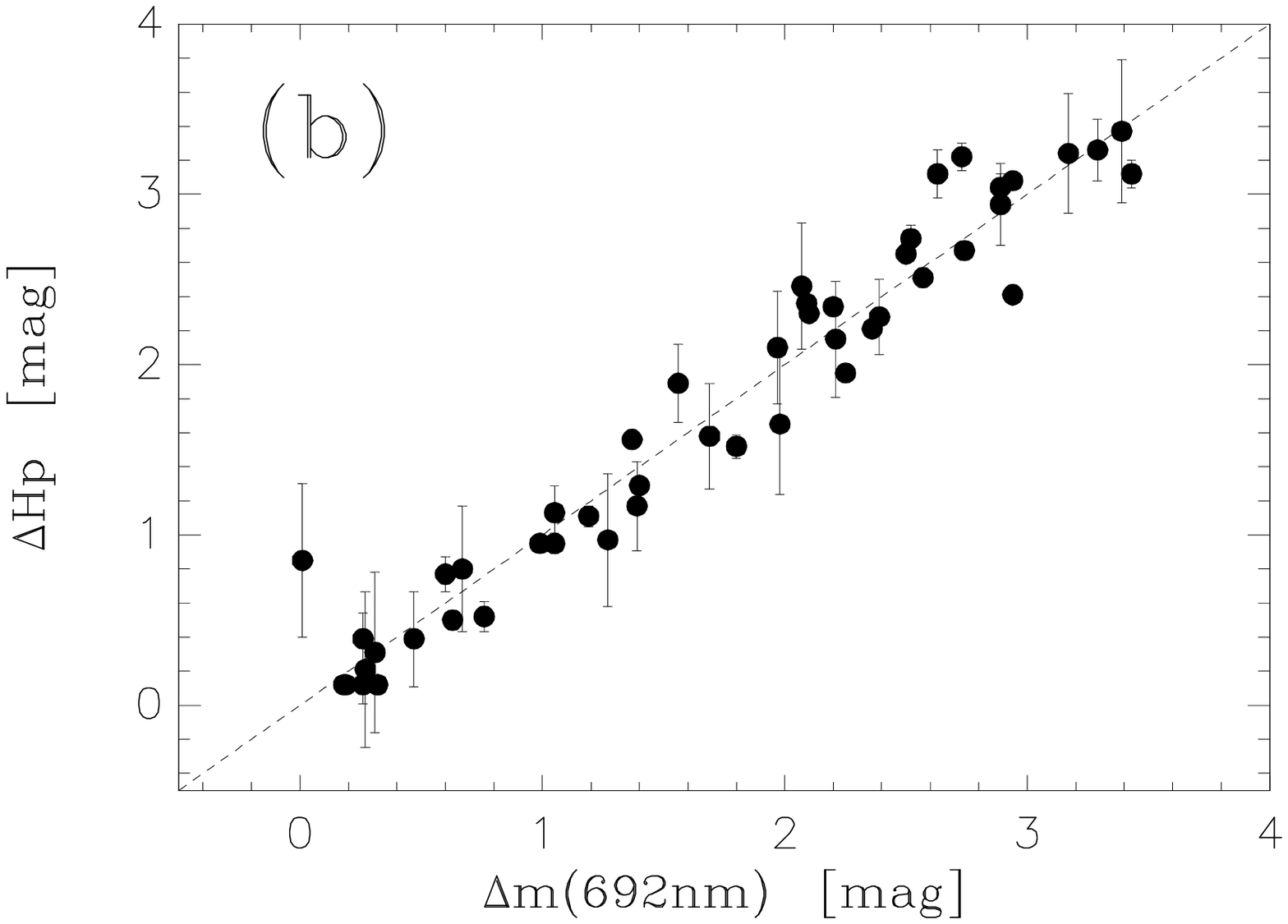}
\caption{ (a) Differences in magnitude difference between the 692-nm measures here and those in the {\it Hipparcos}
Catalogue as a function of seeing times separation. All systems for which the comparison can be made are shown as 
red open circles, and systems with $0.0 < B-V < 0.6$ and the uncertainty in the {\it Hipparcos} $\Delta m < 0.5$ magnitudes
are shown as black filled circles. The error bars are the {\it Hipparcos} uncertainties.
(b) The $\Delta H_{p}$ value appearing in the {\it Hipparcos} 
Catalogue versus the 692-nm magnitude difference in Table 5, for the systems meeting the quality criteria discussed in the text.}
\end{figure}

\clearpage

\begin{figure}[tb]
\plottwo{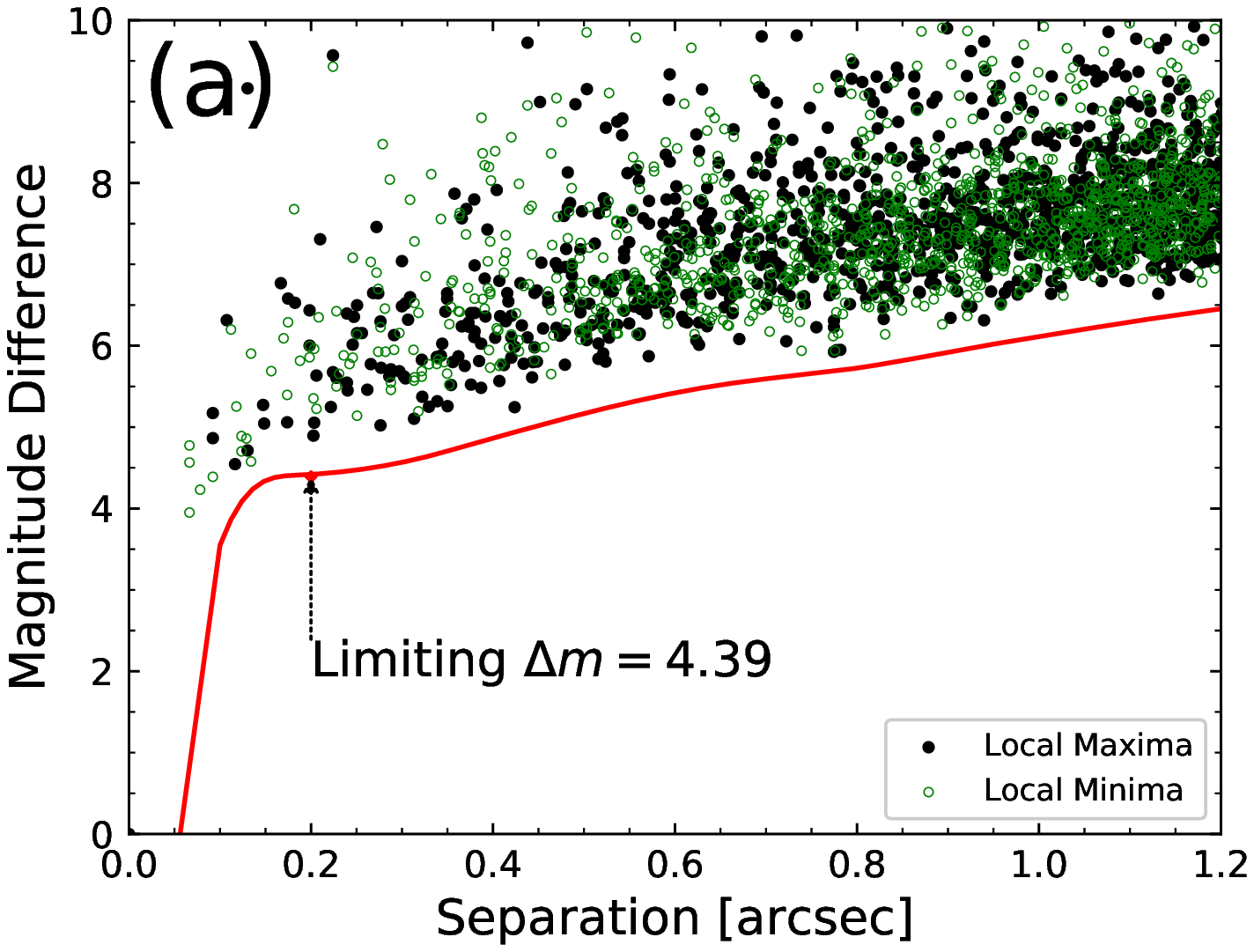}{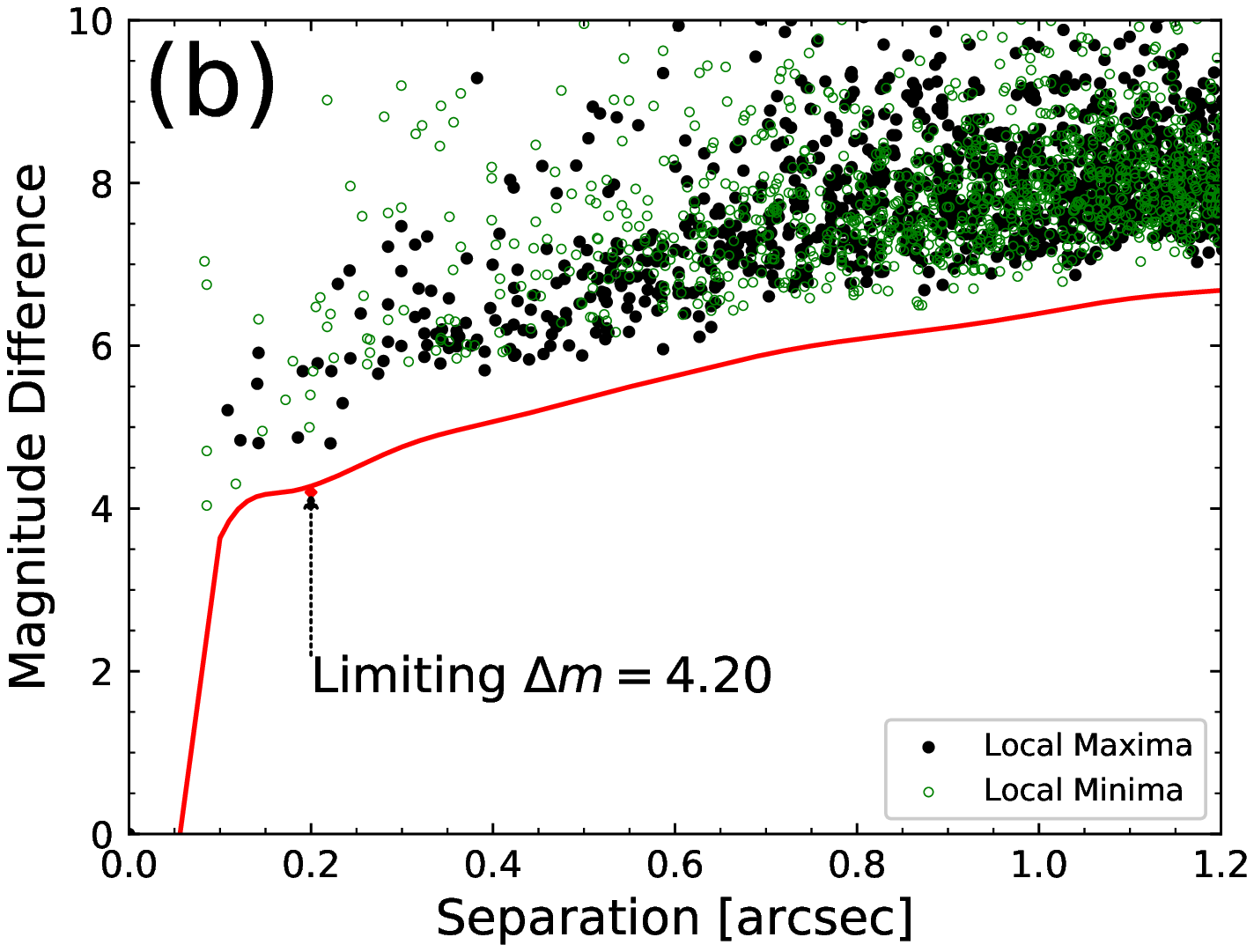}

\vspace{1cm}
\plottwo{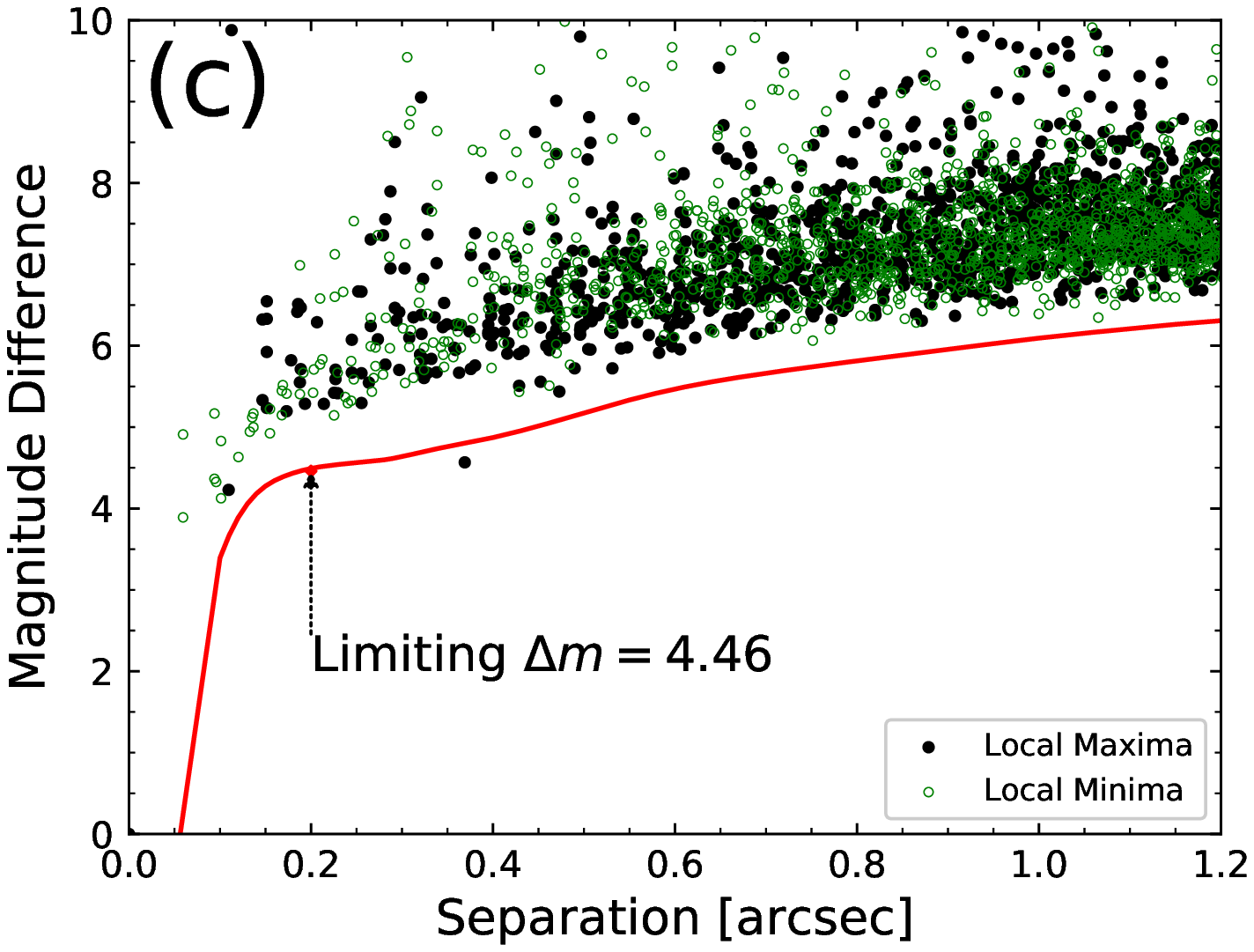}{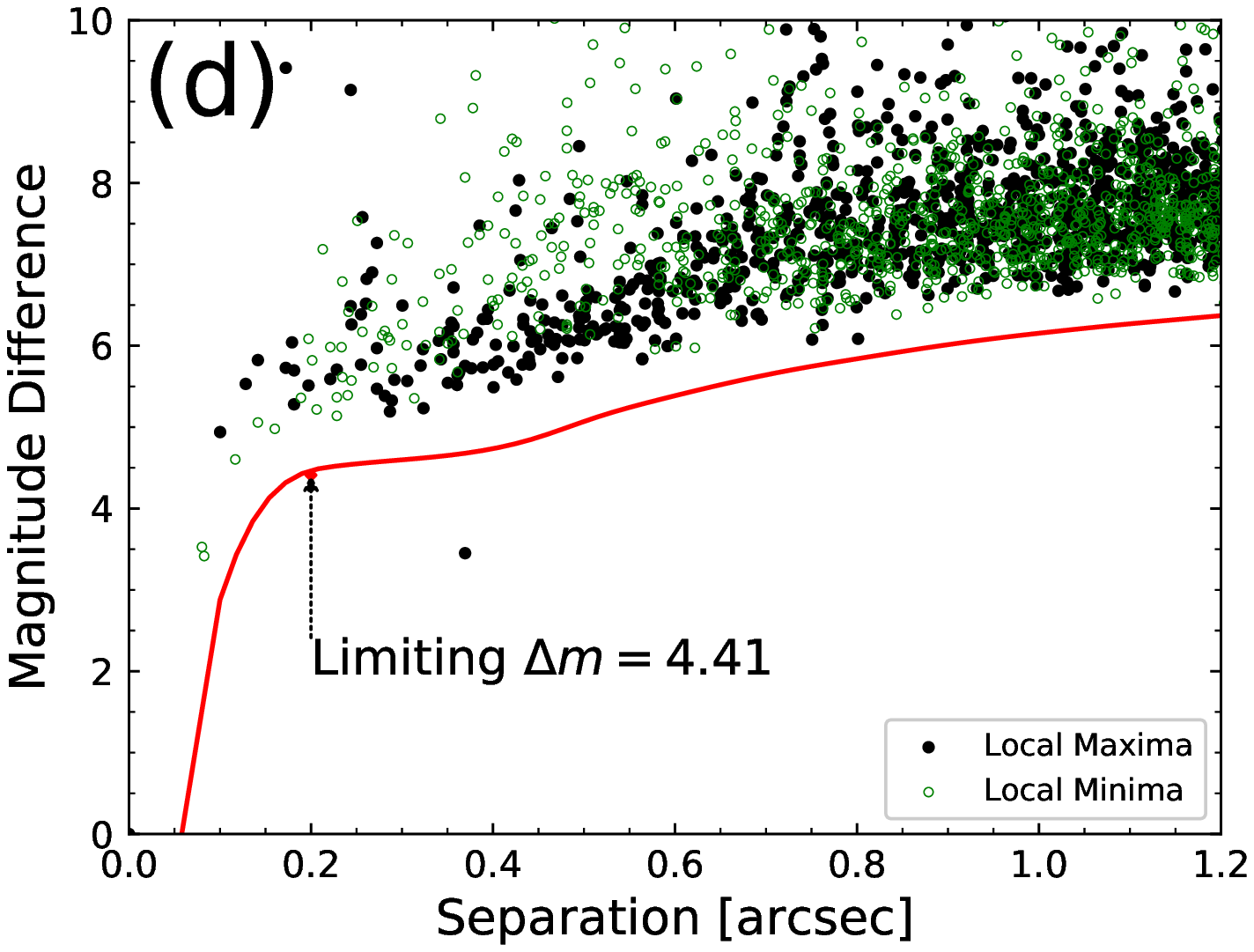}
\caption{ Examples of detection limit curves. (a) V777 Cas = HIP 8980 at 692 nm. (b) V777 Cas at 880 nm. In this case
no companion is detected.
(c) HR 8217 = HIP 105966 at 692 nm. (d) HR 8217 at 880 nm. Here, a faint companion is seen at a separation of 
approximately 0.36 arc seconds from the primary star. This is the first detection of this component, which is listed
as LSC 130 in Table 2. The limiting $\Delta m$ at 0.2 arc seconds from the 
primary star is indicated.}
\end{figure}

\begin{figure}[tb]
\plottwo{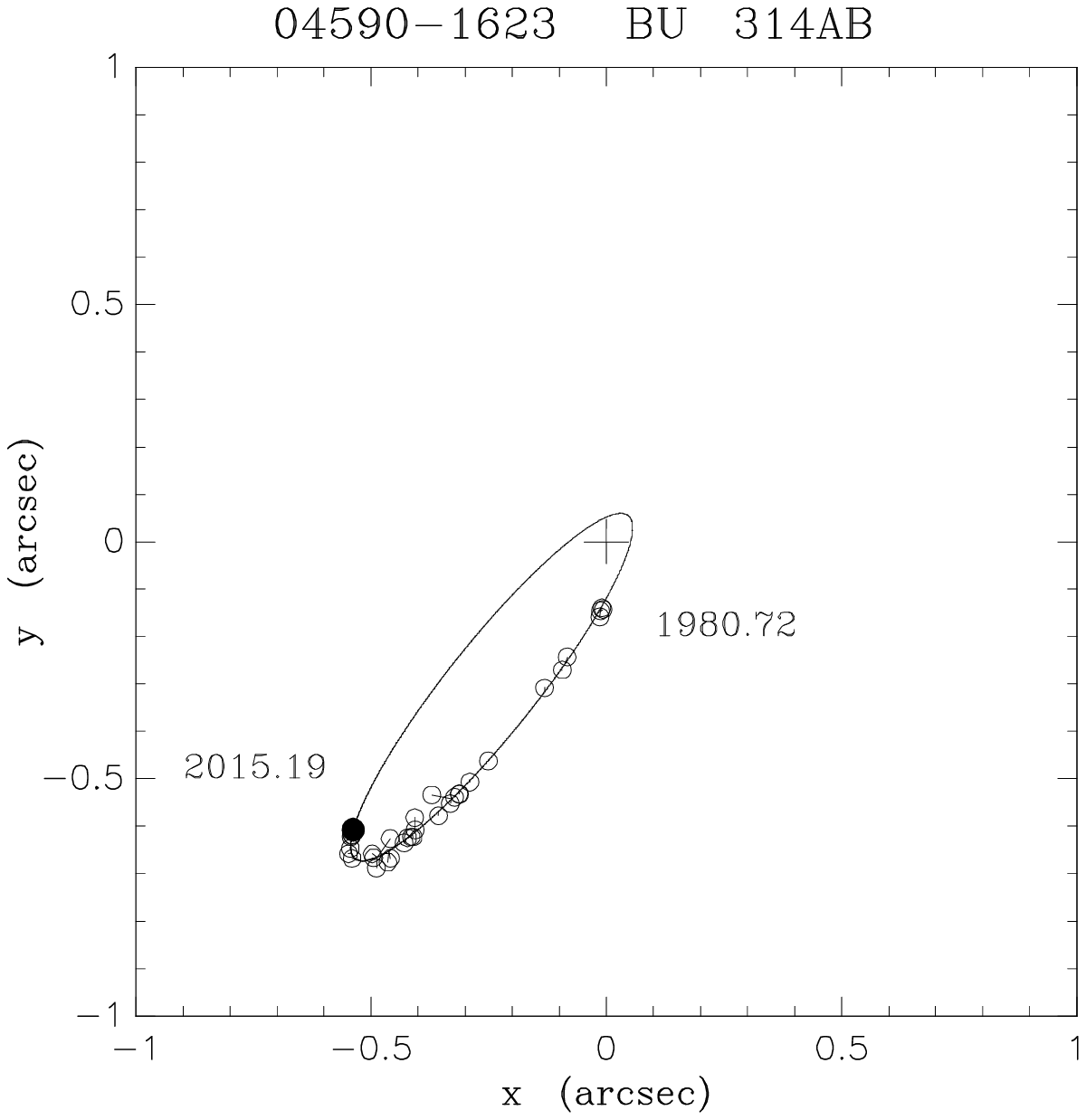}{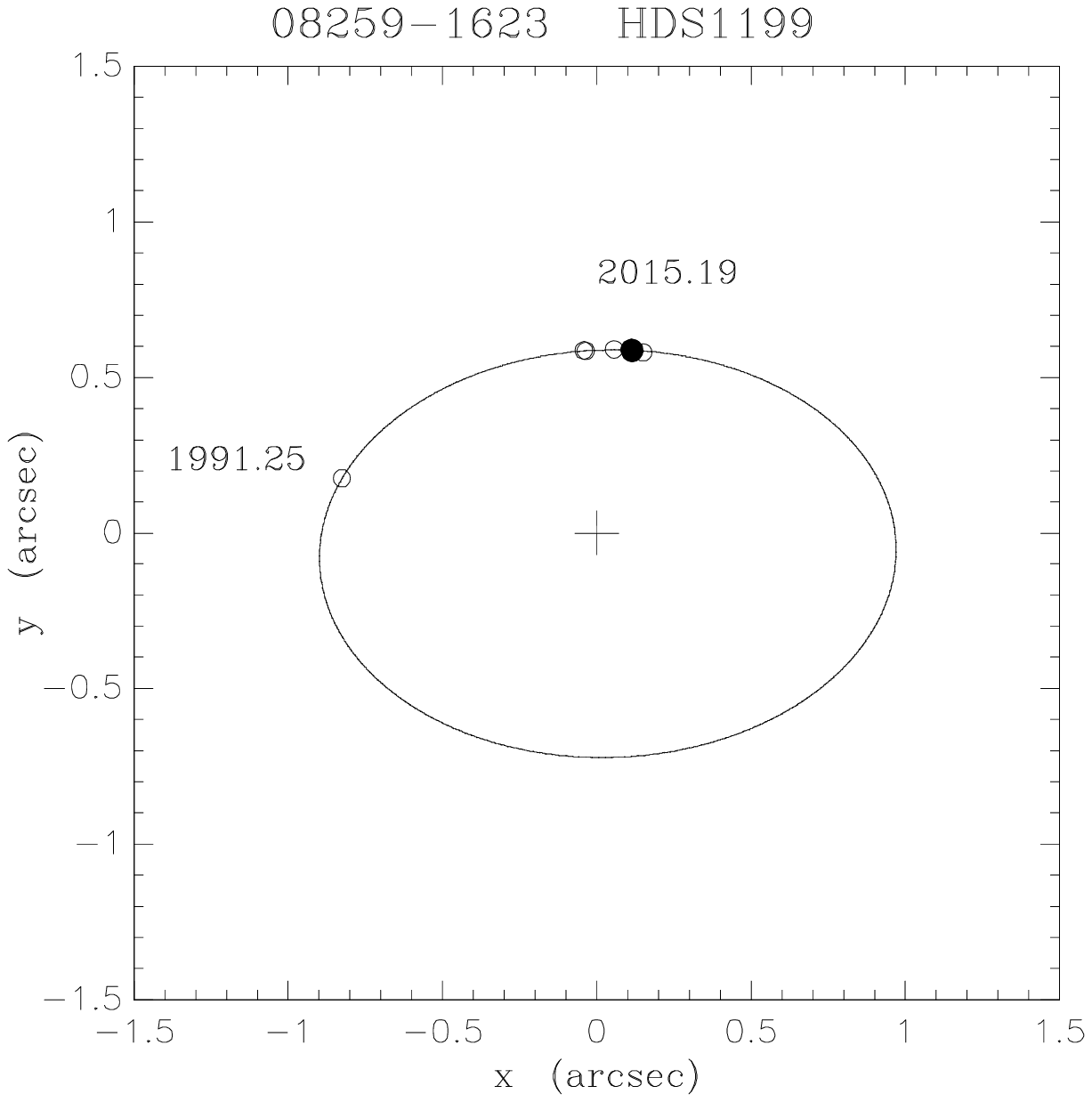}

\vspace{1cm}
\plottwo{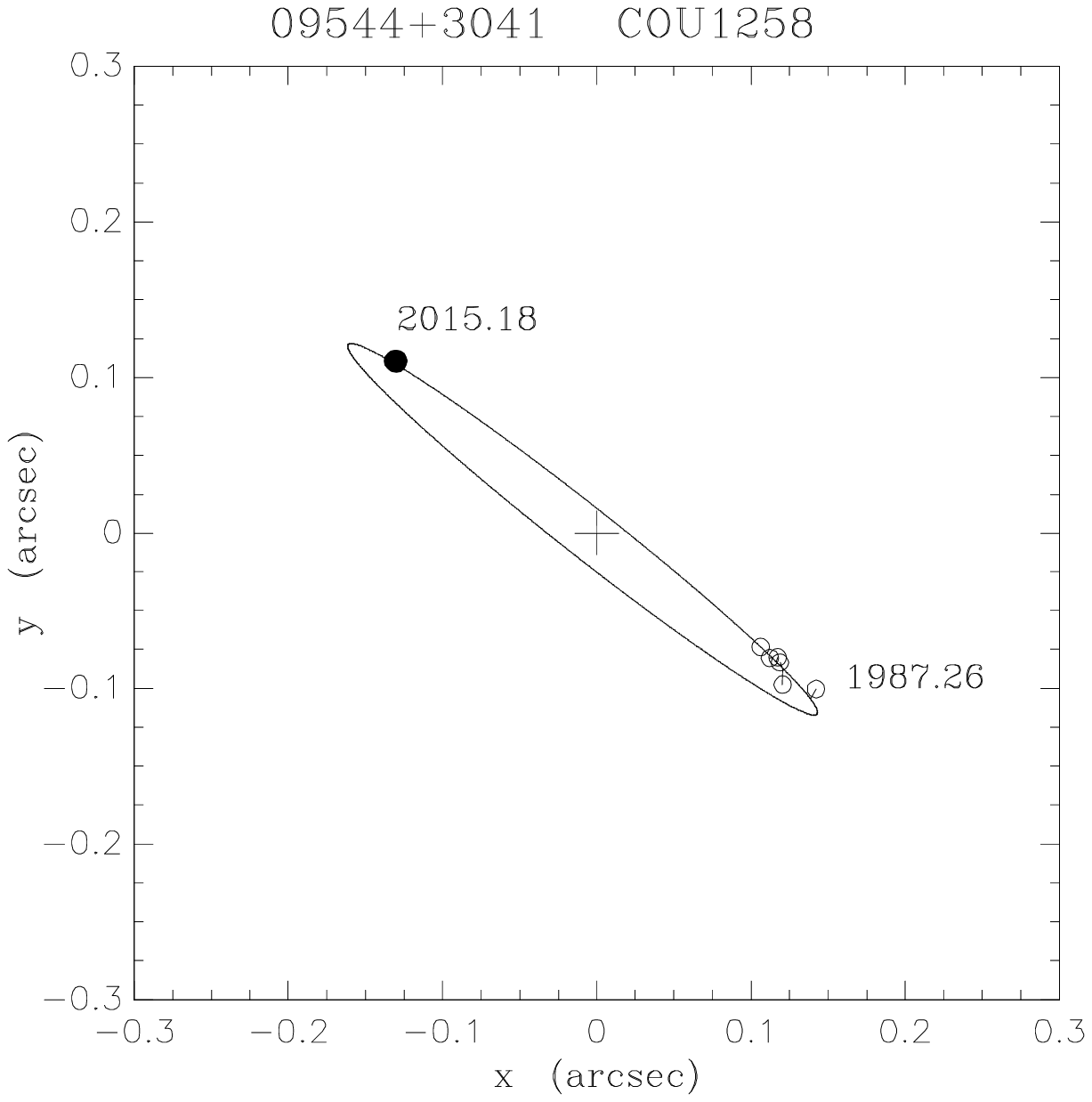}{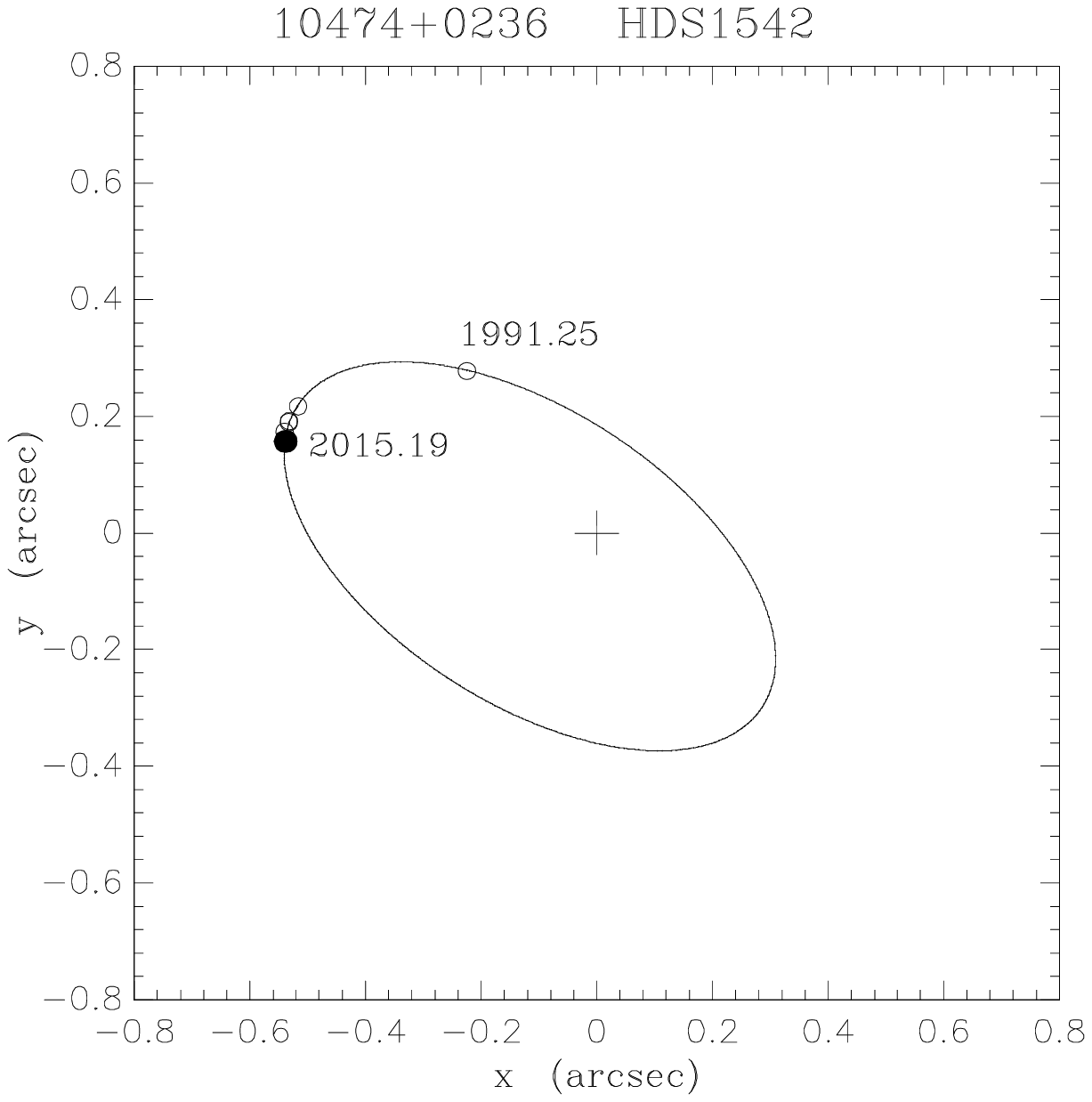}

\vspace{1cm}
\plottwo{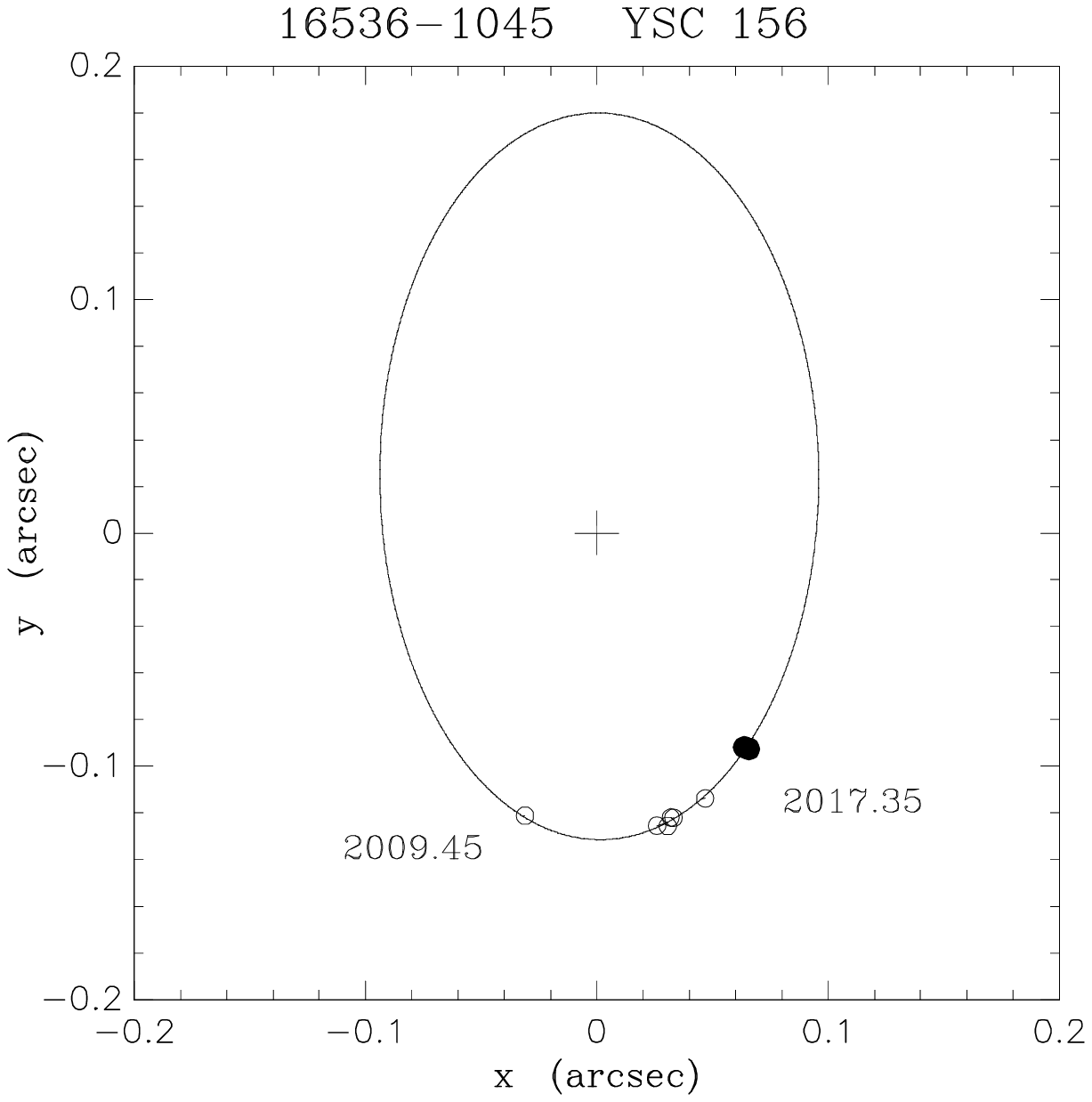}{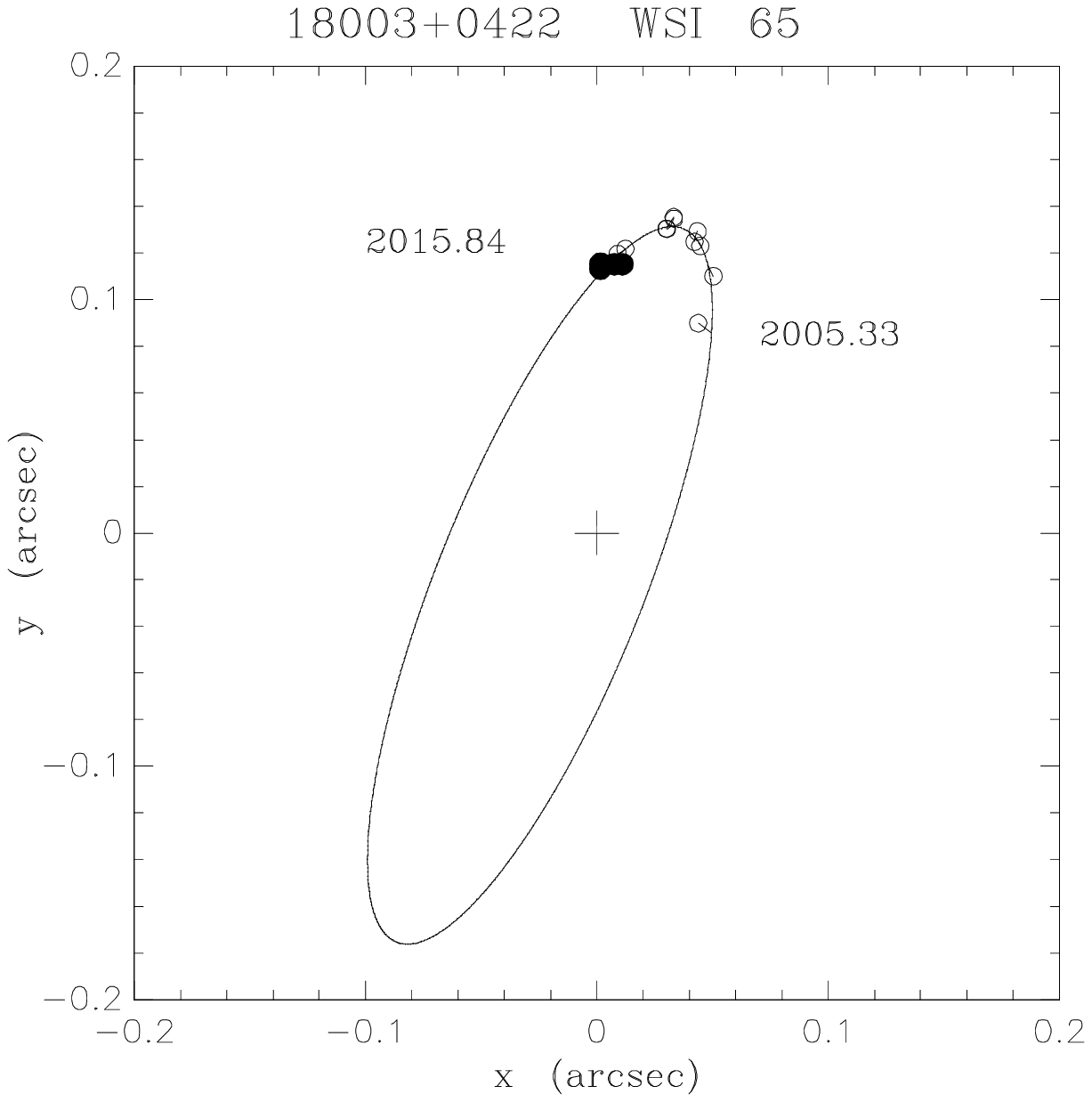}
\caption{
Relative visual orbits for the objects listed in Table 8. The cross marks the position of the primary star, and the relative 
motion of the secondary is shown with the elliptical curve. Data points appearing in the 4th Interferometric Catalog are 
shown as open circles, and measures from the work presented here are shown as filled circles. A line segment joins each
observation to its ephemeris position on the orbits, although many of these are quite small. In all cases, North is down and
East is to the right.}
\end{figure}

\begin{figure}[!htbp]
\figurenum{7}
\epsscale{0.8}
\plotone{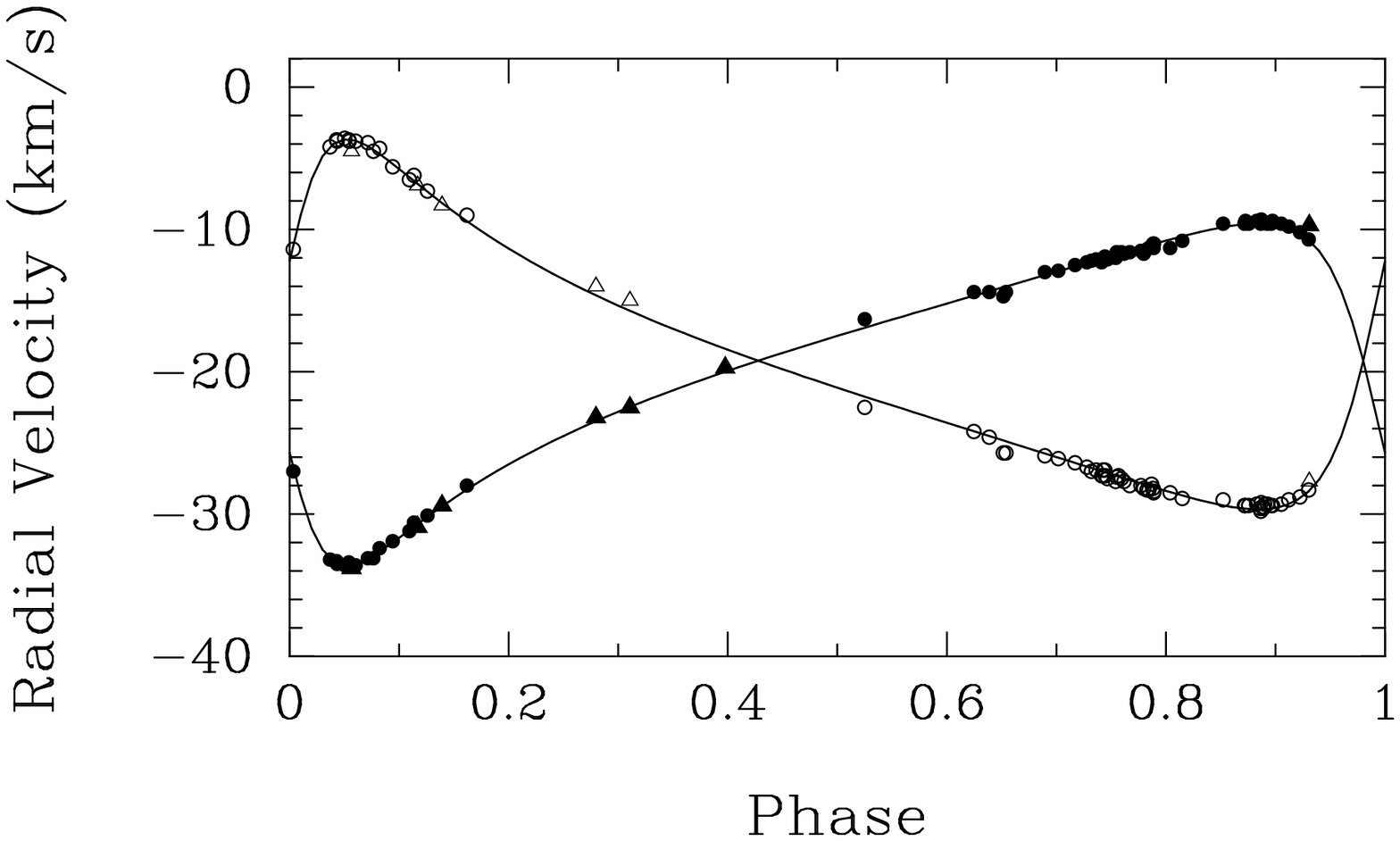}
\figcaption{Radial velocities of HD 22451 = YSC~127 compared with the computed
velocity curves.  Filled and open symbols represent the primary and
secondary, respectively.  Circles = Fairborn Observatory, triangles =
KPNO. Zero phase is a time of periastron passage.
\label{22451sborb}
}
\end{figure}

\begin{figure}[!htbp]
\figurenum{8}
\epsscale{0.8}
\plotone{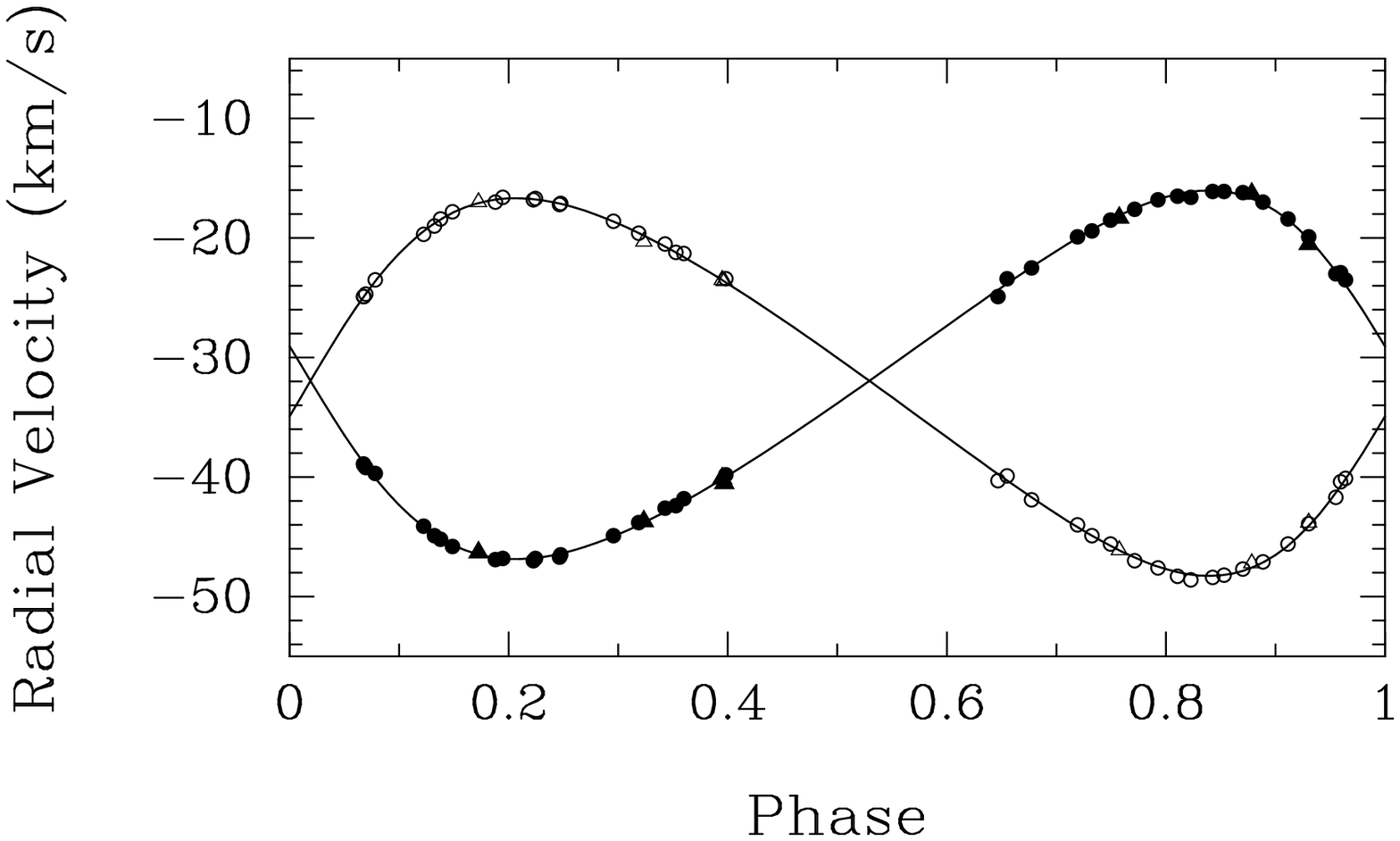}
\figcaption{Radial velocities of HD 185501 = YSC~135 compared with the computed
velocity curves.  Filled and open symbols represent the primary and
secondary, respectively.  Circles = Fairborn Observatory, triangles =
KPNO. Zero phase is a time of periastron passage.
\label{185501sborb}
}
\end{figure}

\begin{figure}[!htbp]
\figurenum{9}
\epsscale{0.8}
\plotone{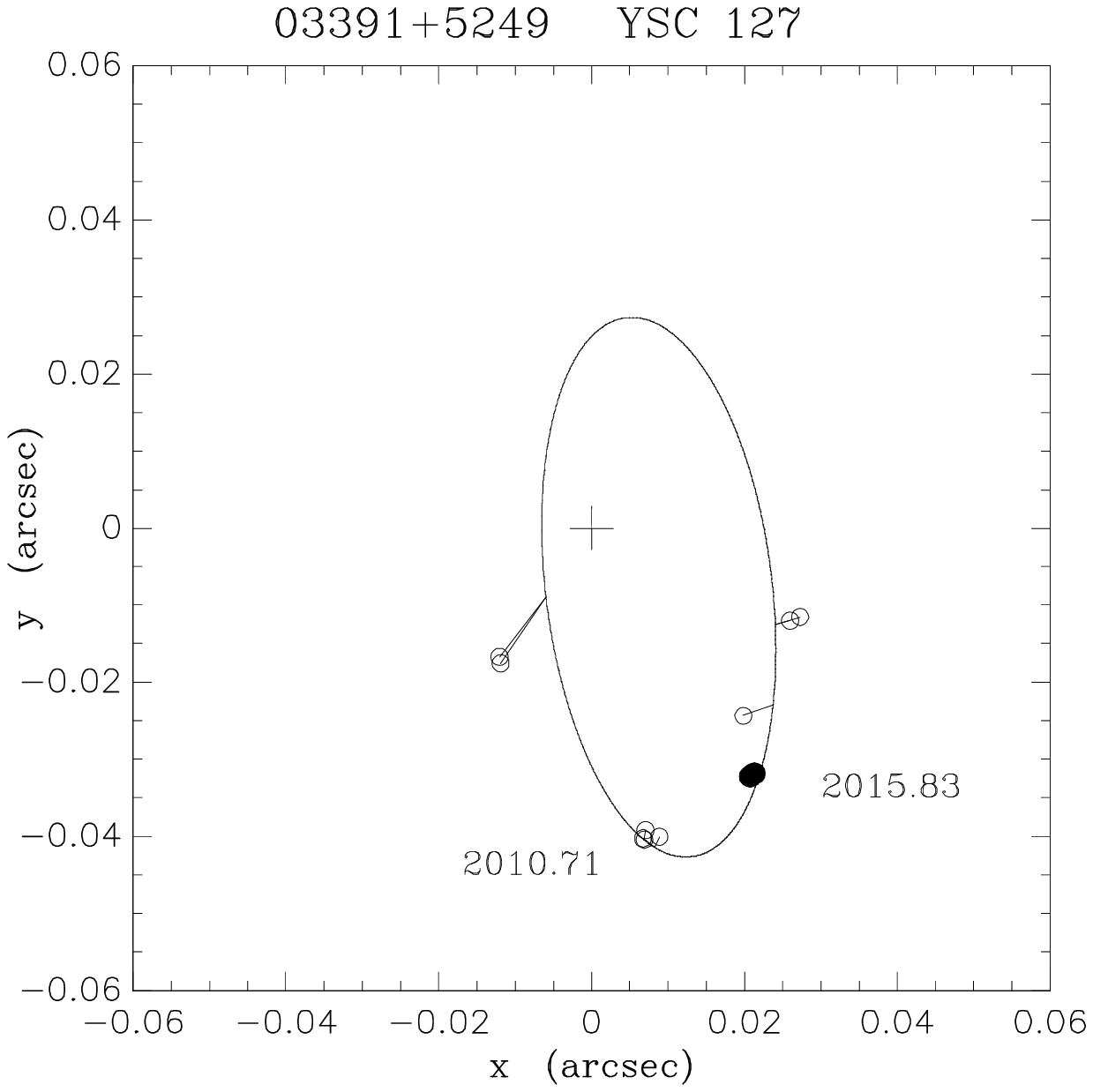}
\figcaption{Visual orbit of HD 22451 = YSC 127 using the joint VB+SB2 orbital elements from Table 9. 
Data points from Table 5 are shown as filled circles
while other data points from the Fourth Interferometric Catalog are shown as open circles.
The cross marks the location of the primary star, and line segments are drawn from the ephemeris position to 
the observed location. North is down, East is to the right. The quadrants have been flipped for data points in Table 5 to be 
consistent with the orbit shown. }
\end{figure}

\begin{figure}[!htbp]
\figurenum{10}
\epsscale{0.8}
\plotone{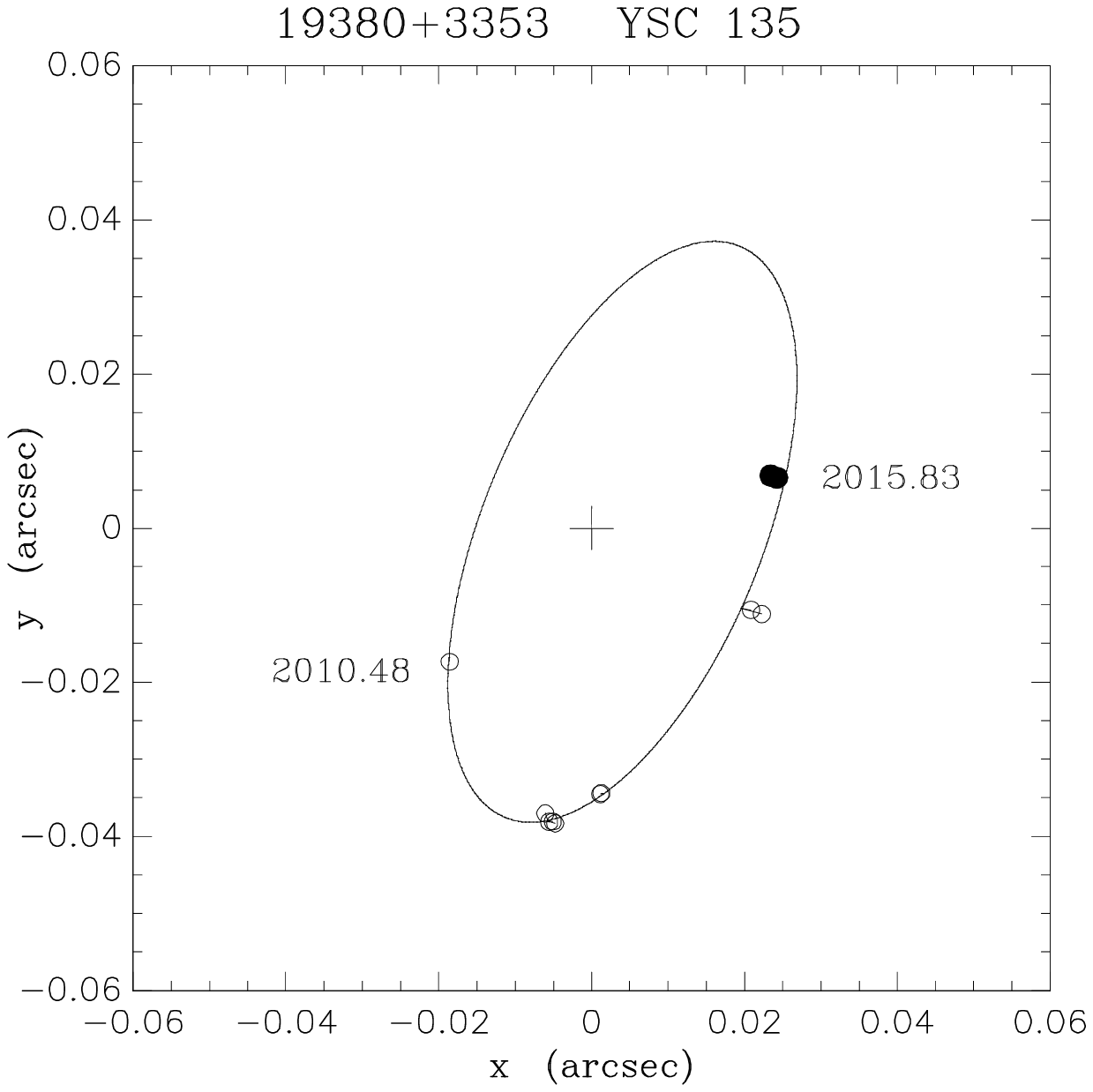}
\figcaption{Visual orbit of HD 185501 = YSC 135 using the joint VB+SB2 orbital elements from Table 9. 
Data points from Table 5 are shown as filled circles
while other data points from the Fourth Interferometric Catalog are shown as open circles.
The cross marks the location of the primary star, and line segments are drawn from the ephemeris position to 
the observed location. North is down, East is to the right.}
\end{figure}

\end{document}